\documentclass[twocolumn,aps,pra,superscriptaddress,10pt]{revtex4-1}
\usepackage{graphicx}
\usepackage{amsmath}
\usepackage{amssymb}
\usepackage{braket}
\usepackage{bm}
\usepackage{dcolumn}
\usepackage{subfigure}
\usepackage{graphicx}
\usepackage[colorlinks=true,linkcolor=blue,citecolor=blue,urlcolor=blue]{hyperref}
\usepackage{soul}
\usepackage{changes}
\usepackage{amsthm,amsmath,amssymb}

\usepackage{mathrsfs}
\usepackage{multirow}

\makeatother

\begin{document}

\title{Laser Manipulation of Spin-Exchange Interaction Between Alkaline-Earth  Atoms in $^1$S$_0$ and $^3$P$_2$ States
}

\author{Shu Yang}

\affiliation{Department of Physics, Renmin University of China, Beijing, 100872,
China}

\author{Yue Chen}

\affiliation{Department of Physics, Renmin University of China, Beijing, 100872,
China}

\author{Peng Zhang}
\email{pengzhang@ruc.edu.cn}

%\selectlanguage{english}%

\affiliation{Department of Physics, Renmin University of China, Beijing, 100872,
China}

\affiliation{Beijing Computational Science Research Center, Beijing, 100193, China}

\date{\today}
\begin{abstract}

Ultracold gases of fermionic alkaline-earth (like) atoms  are hopeful candidates for the quantum simulation of many-body physics induced by magnetic impurities (e.g., the Kondo physics), because there are spin-exchange interactions (SEIs) between two atoms in the electronic ground ($^1$S$_0$) and metastable ($^3$P) state, respectively. Nevertheless, this SEI cannot be tuned via magnetic Feshbach resonance. %So far the only control technique that has been experimentally illustrated is  confinement-induced resonance (CIR). Nevertheless, in many cases the scattering length corresponding to this SEI is much less than the characteristic length of the confinement, and thus the ability of the CIR technique is not significant.
In this work we propose three methods to control the SEI between one atom in the  $^1$S$_0$ state and another atom in the $^3$P$_2$ states or   $^3$P$_2$-$^3$P$_0$ dressed states, with one or two laser beams.
%Explicitly, these two atoms experience a zero-range effective  interaction
%proportional to $[\!\frac{A_{x}}2\hat{\sigma}_x^{(S)}\hat{\sigma}_x^{(P)}\!
%+\!\frac{A_{y}}2\hat{\sigma}_y^{(S)}\hat{\sigma}_y^{(P)}\!
%+\!\frac{A_{z}}2\hat{\sigma}_z^{(S)}\hat{\sigma}_z^{(P)}\!+\!A_0]$, where ${\hat{\sigma}}_{x,y,z}^{(S(P))}$ are
%the Pauli operators for the pseudo-spin of the atom in the $^1$S$_0$ ($^3$P) state, and using our approaches one can manipulate the parameters $A_{x,y,z,0}$  by changing the intensities of the laser beams.
These methods are based on the spin-dependent AC-Stark shifts of the $^3$P$_2$ states, or the $^3$P$_2$-$^3$P$_0$ Raman coupling.  We show that due to the structure of alkaline-earth (like) atoms,  the  heating effects  induced by the  laser beams of our methods are very weak. For instance, for  ultracold Yb atoms,  AC-Stark-shift difference of variant spin states of the $^3$P$_2(F=3/2)$  level, or the strength of the $^3$P$_2$-$^3$P$_0$ Raman coupling, could be of the order of $(2\pi)$MHz, while the heating rate (photon scattering rate) is only of the order of Hz. As a result, the Feshbach resonances, with which one can efficiently control the SEI by changing the laser intensity, may be induced by the laser beams with low-enough heating rate, even if the scattering lengths of the bare inter-atomic interaction are so small that being comparable with the length
scale associated with the van der Waals interaction.
\end{abstract}
\maketitle
%\end{CJK*}

\section{ Introduction}
\label{introduction}

In recent years, the ultracold gases of fermionic  alkaline-earth (like) atoms have attracted much attention \cite{He2019,Yoshiro2020}. One important motivation for studying  this system is that there are spin-exchange interactions (SEIs) between two fermionic alkaline-earth (like) atoms ({\it e.g.,} two $^{173}$Yb atoms or two $^{171}$Yb atoms) in the electronic $^1$S$_0$ and $^3$P states, respectively, which plays a central role on the quantum simulation of many-body models with magnetic impurities ({\it e.g}, the Kondo or Kondo-lattice models)
\cite{Scazza2014,Hofer2015,Cappellini2019,DaWu2022,Ono2019,Abeln2021,Yejun2014,Yejun2010,Cheng2022,Rey2010,Yoshiro2021,Cappellini2014,Norio2015,Danshita2019,Nishida2013,
Demler2013,Gyu-Boong2016,Gyu-Boong2018,Riegger2018,Ren2020,ShiTao2018,Ren2016,Takahashi2021}.
Explicitly, the atoms in the  $^3$P state can be individually confined in the sites of a deep optical lattice, and play the role of the local magnetic impurities, so that the two-body loss induced by the collision of two $^3$P atoms can be avoided. In addition, the moving $^1$S$_0$ atoms can play the role of the itinerant electrons in the Kondo-type models.
To realize this quantum simulation, it is important to develop the techniques to manipulate the SEI
\cite{Ren2020,Ren2018,Ren2016,Cheng2017,Peng2020,Riegger2018,ShiTao2018,Gyu-Boong2018,Avishai2018}.

To avoid the heating loss induced by the spin-exchange process, the difference between the Zeeman energies of the atoms
before and after this process should be lower than the temperature of the ultracold gases. Thus, the control of the SEI should be done under zero or low-enough magnetic field, and thus is difficult to be realized via magnetic Feshbach resonance
\cite{Ren2015}.
Due to this fact,  people studied the manipulation of SE interaction via a confinement-induced resonance (CIR) \cite{Olshanii1998} under zero magnetic field
\cite{Ren2018,Ren2016,Cheng2017,Peng2020}.
This technique has been experimentally realized for the control of the nuclear SE interaction between ultracold  $^{173}$Yb atoms
\cite{Riegger2018}.

Nevertheless, mostly the CIR occurs only when the inter-atomic scattering length in the three-dimensional (3D) free space is comparable with the characteristic length of the confinement, which is usually of the order of $1000a_0$. For current experiments of ultracold
alkaline-earth (like) atoms,
 this condition is only partly satisfied by   $^{173}$Yb atoms, for which one of the two scattering lengths related to the SEI is about $2000a_0$ \cite{Hofer2015,Cappellini2019,DaWu2022}. For other systems, {\it e.g.,} $^{171}$Yb atoms, the relevant 3D scattering lengths are of the order of $100a_0$ \cite{Oscar2003,Ono2019,Abeln2021, DaWu2022}, i.e., much less than the confinement characteristic length,
and thus the control effect of the CIR approach to be weak.
On the other hand, the interaction between atoms in $^1$S$_0$ and $^3$P$_0$ states  includes not only the SEI but also a spin-independent term.
In current experiments of  $^{173}$Yb or $^{171}$Yb atoms,
this term is very strong, so that the spin-exchange effects may be suppressed \cite{Cheng2022}.
 Therefore, it would be helpful if more control techniques
for the SEI between  alkaline-earth (like) atoms
 can be developed.

In this work, we propose three methods for  controlling the SEI  between two fermionic  alkaline-earth (like) atoms with pseudo-spin $1/2$ via one  or two  laser beams. Explicitly,
 one atom is in the $^1$S$_0$ state and another one in the $^3$P$_2$ state (methods I and II) or a $^3$P$_2$ -$^3$P$_0$ dressed state (method III) \cite{1page,2page}. So far the SEI of alkaline-earth like atoms has been only observed with atoms being in $^1$S$_0$ and
$^3$P$_0$ states. Nevertheless, for our systems with pseudo-spin 1/2 atoms in the $^1$S$_0$ and
$^3$P$_2$ states, there  also exits
SEI ({\it i.e.}, the exchange of the pseudo-spin states) processes. These processes
are induced by a similar mechanism
as the one for the $^1$S$_0$ and
$^3$P$_0$ atoms,
 and are
 permitted by the selection rule of the corresponding inter-atomic interaction potential, as shown below.
 In this work, we consider the $^3$P$_2$ states because for these states the laser-induced effects which can be used for the manipulation of SEI ({\it e.g.}, the spin-dependent AC-Stark effects) are much more significant than the ones of $^3$P$_0$ states.

Our approaches are based on the spin-dependent AC-Stark shifts of   $^3$P$_2$ states (methods I and II), or the laser-induced Raman coupling between $^3$P$_2$ and $^3$P$_0$ states (method III). Explicitly, for the systems of these methods, there are both open and closed channels of
the  spin-exchange scattering processes, and the
energy gap between the open and closed channels
just equals to (or has the same order of magnitude with)
the AC-Stark shift difference $\Delta_{\rm AC}$  between $^3$P$_2$ states with different magnetic quantum numbers, or the effective Rabi frequency $\Omega_{\rm eff}$ of the $^3$P$_2$-$^3$P$_0$ Raman coupling. Therefore, one can control the amplitude of the spin-exchange scattering of the atoms incident from the open channel, or the effective inter-atomic SEI, by tuning $\Delta_{\rm AC}$ or $\Omega_{\rm eff}$ via changing the laser intensity.

More importantly, we show that the heating effects induced by the laser beams are quite weak. This is due to the structure of alkaline-earth (like) atoms, and
is very different from the situations of the  ultracold alkaline atoms under similar laser manipulations ({\it e.g.}, the Raman coupling between different hyperfine states of electronic ground state), where the lasers mostly induce strong heating.
  For instance, for a Yb atom the heating rate (photon scattering rate) could be just of the order of Hz when $\Delta_{\rm AC}$ or $\Omega_{\rm eff}$, and thus the energy gap between the open and closed channels, is of the order of $(2\pi)$MHz, and is comparable with the van der Waals energy scale $E_{\rm vdW}$.
  On the other hand,  the potentials of the closed channels are very possible to support $s$-wave bound states with the binding energies $|E_b|$ being comparable with or less than the van der Waals energy $E_{\rm vdW}$, even in the absence of a $s$-wave resonance.
%if  the scattering length of the bare inter-atomic interaction is as small as the length scale $\beta_6$ associated with the van der Waals interaction
For instance, for the cases of a single-channel  van der Waals interaction
we have $|E_b|<0.99
E_{\rm vdW}$ for $a_s>\beta_6$ \cite{Bo2004}, with $\beta_6$ being the
 length scale  associated with this van der Waals interaction \cite{footnote1}.
 Moreover, for the systems of our methods II and III, the $s$-wave states of the open and closed channels are coupled to each other. Thus, for these systems, no matter if the closed channels are on resonance,
it is always very possible that one can
make the threshold of the open channel  be resonant to a closed-channel bound state  by tuning the laser intensity, and thus
 induce Feshbach resonances for these two atoms, while keeping the heating rate low enough. Using these resonances one can efficiently manipulate the effective SEI.

For the systems of our method I where the $s$-wave states of the open channels are only coupled to the $d$-wave closed-channel bound states, the above kind of ``low-heating" Feshbach resonance occurs when the closed channels are close to a $d$-wave resonance. Nevertheless,  for our system there are four degenerate closed channels which are coupled with each other. Therefore, the probability for the appearance of these resonances is much larger than the one of a single-channel van der Waals potential.

For the systems of all the methods I, II, and III, we can always treat the atoms in the $^1$S$_0$ states and the relevant $^3$P states as  two distinguishable atoms with pseudo-spin $1/2$. The effective Hamiltonian of these two atoms can be expressed as
\begin{eqnarray}
H_{\rm 2body}^{\rm (eff)}=\frac{{\bm p}_S^2}{2m}+\frac{{\bm p}_P^2}{2m}+{\hat V}_{\rm eff}, \nonumber
\end{eqnarray}
where
$m$ is the single-atom mass, ${\bm p}_{S(P)}$ is the momentum operator of the atom in the $^1$S$_0$ ($^3$P) state. Here
 the effective inter-atomic interaction ${\hat V}_{\rm eff}$ is given by {$(\hbar=1)$:}
\begin{eqnarray}
&&{\hat V}_{\rm eff}=\frac{2\pi}{\mu}\times\nonumber\\
&&\left[\!\frac{A_{x}}2\hat{\sigma}_x^{(S)}\hat{\sigma}_x^{(P)}\!
+\!\frac{A_{y}}2\hat{\sigma}_y^{(S)}\hat{\sigma}_y^{(P)}\!
+\!\frac{A_{z}}2\hat{\sigma}_z^{(S)}\hat{\sigma}_z^{(P)}\!+\!A_0
\right]\!
\delta({\bf r})\frac{\partial}{\partial r}(r\cdot),\nonumber\\
\label{11}
\end{eqnarray}
with $\mu\equiv m/2$ and ${\bf r}$ being reduced mass and the inter-atomic position, respectively, and
${\hat{\sigma}}_{x,y,z}^{(S(P))}$ being
the Pauli operators for the pseudo-spin of the atom in the $^1$S$_0$ ($^3$P) state. Namely, the pseudo-spin states of the two atoms are {\it degenerate} eigen-states of the effective two-atom  free Hamiltonian ${{\bm p}_S^2}/{(2m)}+{{\bm p}_P^2}/{(2m)}$. In addition, the
effective interaction $\hat{V}_{\rm eff}$ is described by the parameters $A_{x,y,z,0}$. For instance, the strength of the effective SEI is $(A_x+A_y)/2$, and the strength of the spin-independent interaction is $A_0$.
In addition, for the systems of methods I and III we always have $A_x=A_y$, while for the system of method II $A_x$ and $A_y$ may be unequal.
Using our methods one can resonantly control the parameters $A_{x,y,z,0}$ via changing the laser intensity.

Since we  lack the detailed parameters for the bare interaction potential between atoms in $^1$S$_0$ and $^3$P$_2$ states,
so far we cannot perform quantitatively accurate calculations for experimental systems.
Therefore, in this work we qualitatively  illustrate the three methods with two-body calculations for a multi-channel square-well interaction model. Our results show that    the effective interaction can be tuned to be either ``anti-ferromagnetic-like"  or ``ferromagnetic-like" for many cases,
where the lowest eigen state of the pseudo-spin operator in the square  bracket of Eq.~(\ref{11}) is the singlet state or the pseudo-spin-polarized states, respectively. In addition,
the absolute values $|A_{x,y,z,0}|$ can be controlled in a broad region  ({\it e.g.,} from zero to $1000a_0$). One can also completely ``turn off" the spin-independent interaction ({\it i.e.,} tune $A_0$ to be zero) while keeping the SEI strength $(A_{x}+A_y)/2$ to be finite.
%The effects of these methods for realistic systems,  especially the position and width of the Feshbach resonances as well as the corresponding laser-induced heating rate, are to be examined via experiments or coupled-channel calculations with detailed information of inter-atomic interaction.

%Nevertheless, here we also emphasis that in our calculation we do find some cases  where the resonances do not occur when  $\Delta_{\rm AC}$  of method I or  $\Omega_{\rm eff}$ of method II is less than $(2\pi)10$MHz. These cases are relatively easier to appear for method I, and  are consistent with the fact that sometimes the binding energy of the shallowest bound state of a van der Waals interaction potential can be larger than  $E_{\rm vdW}$ by one order of magnitude. The effects of these methods for realistic systems is to be examined via experiments or coupled-channel calculations with detail informations of inter-atomic interaction. (Notice that  our system is complicated because there are many degenerate closed channels which are coupled with each other, as shown below. Thus, for a specific type of atom one cannot predict if  resonances can appear for a certain region of $\Delta_{\rm AC}$   or  $\Omega_{\rm eff}$ ({\it e.g.},  $\Delta_{\rm AC}<(2\pi)10$MHz   or  $\Omega_{\rm eff}<(2\pi)10$MHz), even with the analysis based on the theory of single-channel van der Waals potential.)

The remainder of this paper is organized as follows. In Sec.~\ref{m1},~\ref{nm2}, and ~\ref{m2} we show the principles of methods I, II, and III, respectively, and illustrate the control effects via the square-well model.
In Sec.~\ref{summary} we provide some discussions, including a comparison  of the advantages and disadvantages of these three methods. Some details of our calculations are given
in the appendixes.

\section{ Method I}
\label{m1}

In this and the following two sections,  we introduce our methods for the manipulation of SEI in detail. For clearance,  we take the system of two $^{171}$Yb($I=1/2$) atoms as an example. The generalization of our methods for atoms of other species is straightforward.

Our method I is based on the strong
spin-dependent AC-Stark effect of $^3$P$_2$ states. In the following, we first introduce this effect and then show how to use this effect to  control the SEI.

\begin{figure}
\includegraphics[width=0.48\textwidth]{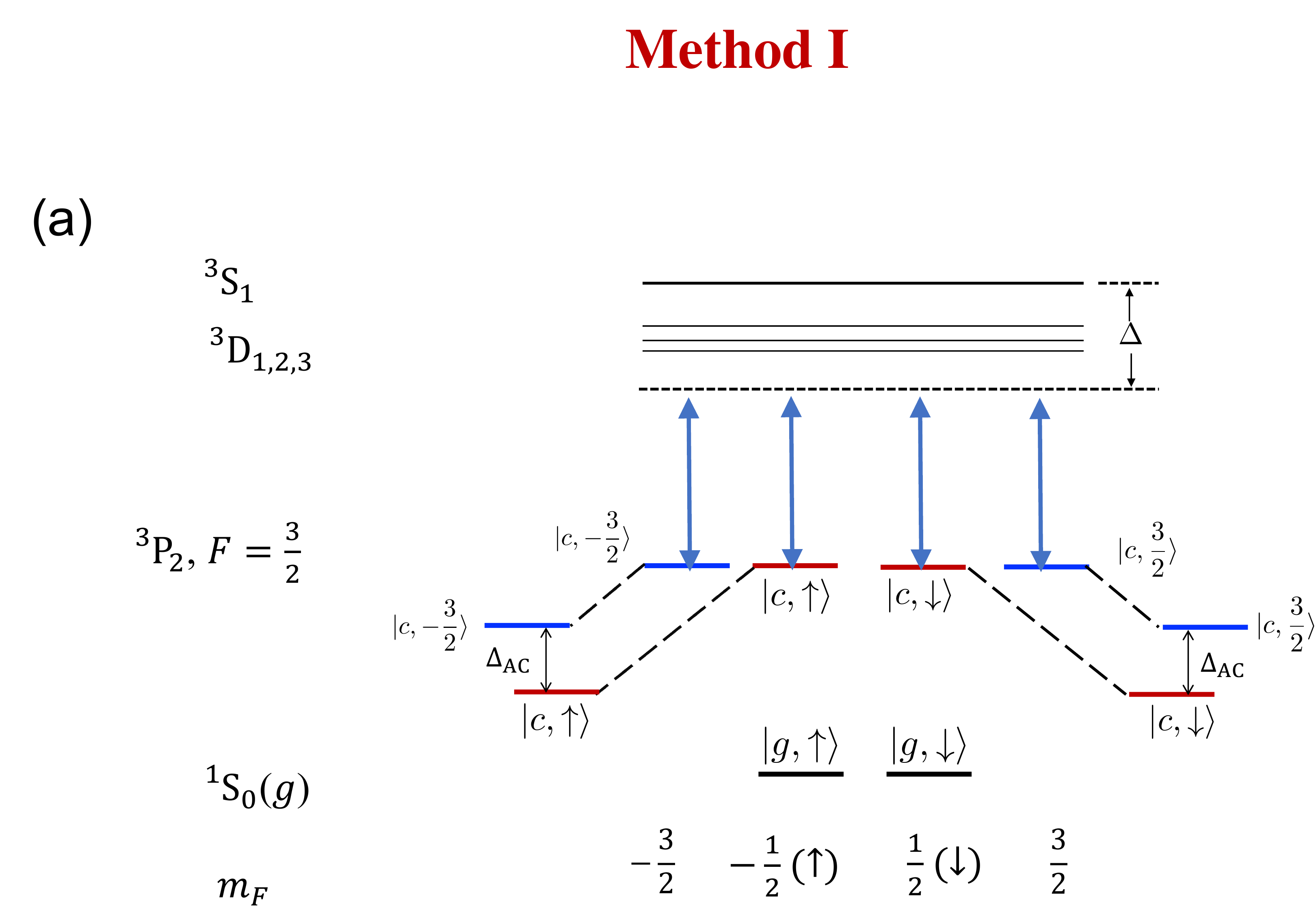}\\
\vspace{1cm}
\includegraphics[width=0.48\textwidth]{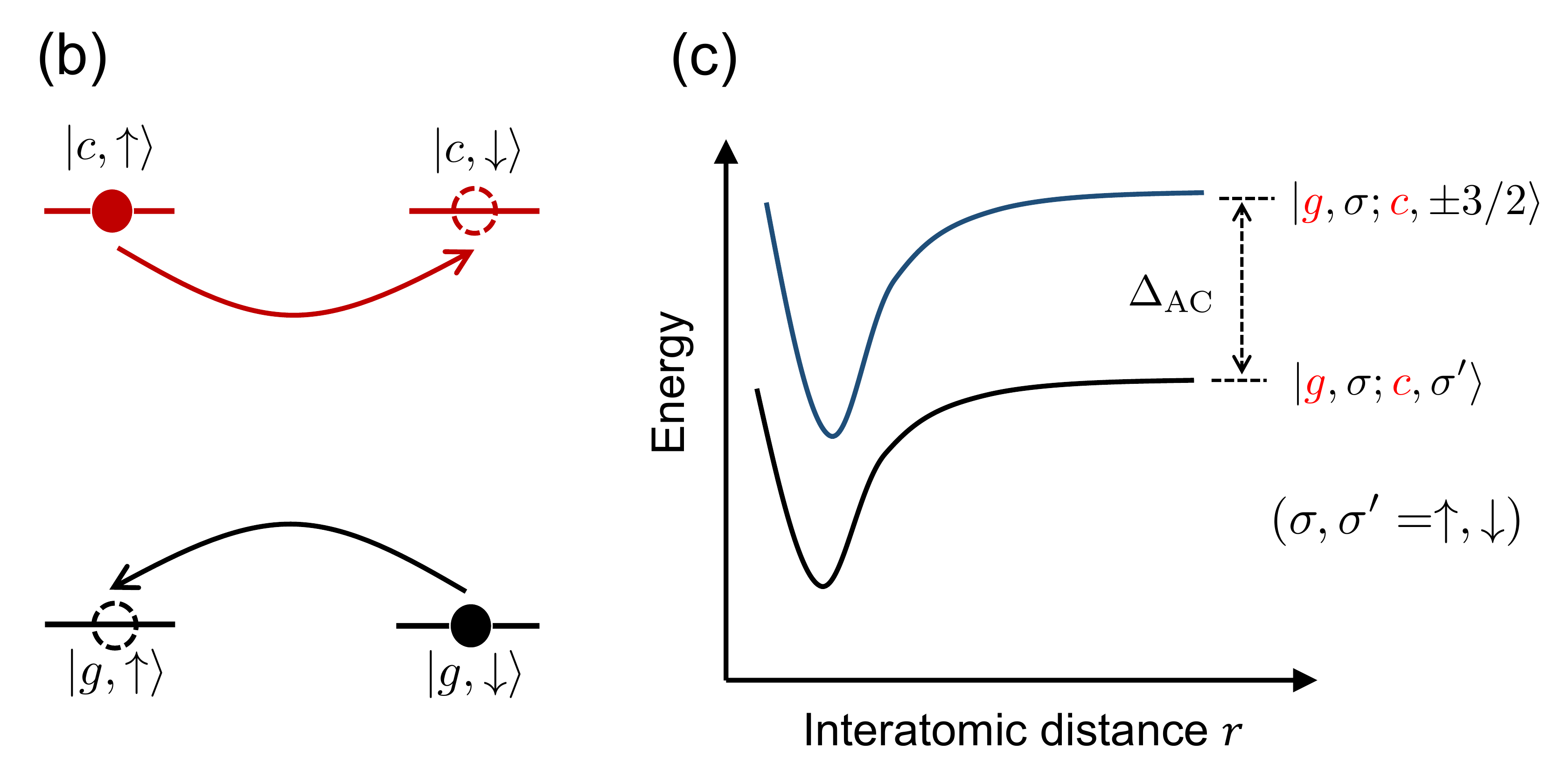} 
\caption{(color  online) Schematic illustration of method I. {\bf  (a):} Single-atom energy levels and the $\pi$-polarized laser beam (blue lines with arrows). We also show the $^3$P$_2$ levels shifted by
the beam.
 {\bf  (b):} A Two-atom spin-exchange process. The black (red) filled and dashed circles represent the $g(c)-$atom before and after a collision, respectively. Both this process and the inverse one are studied in this work. {\bf  (c):} The inter-atomic scattering channels. Here the solid curves represent the potentials of each channel. The
 coupling potentials between different channels are not shown in the figure.}
\label{scheme1}
\end{figure}

\subsection{AC-Stark Shifts and Heating Effects of $^3$P$_2$ States}
\label{m1a}

As shown in Fig.~\ref{scheme1}(a), in our scheme a $\pi$-polarized laser beam is applied  at a zero magnetic field,
so that the $^3$P$_2 (F=3/2)$ states are far-off resonantly coupled
to the excited states ({\it e.g.}, the $^3$S$_1$ and $^3$D$_{1,2,3}$ states).
Explicitly,  all of the detunings of this  beam with respect to the
transition to the excited states \cite{3page} are much larger than the fine splitting of these states.
As a result, the energies of the $^3$P$_2 (F=3/2)$ states  are  shifted via the AC-Stark effect.
We denote the
$^3$P$_2 (F=3/2)$ states with $m_F=-1/2 (+1/2)$ and   $m_F=\pm 3/2$ as $|c,\uparrow\!(\downarrow)\rangle$ and $|c,\pm 3/2\rangle$, respectively.
 It is clear that  AC-Stark shifts $E^{\rm (AC)}_{\xi}$of state $|c,\xi\rangle$  ($\xi=\uparrow,\downarrow,\pm 3/2$)  satisfies
\begin{eqnarray}
E^{\rm (AC)}_{\uparrow}&=&E^{\rm (AC)}_{\downarrow};\quad E^{\rm (AC)}_{-3/2}=E^{\rm (AC)}_{+3/2}.
\end{eqnarray}
We further define the difference between the AC-Stark shifts of states $|c,\uparrow\!(\downarrow)\rangle$ and $|c,\pm 3/2\rangle$  as (Fig.~\ref{scheme1}(a)):
\begin{eqnarray}
\Delta_{\rm AC}&\equiv&E^{\rm (AC)}_{-3/2}-E^{\rm (AC)}_{\uparrow}.\label{dac}
\end{eqnarray}

Here we emphasize that,
the spin-dependence of the AC-Stark effect for the $^3$P$_2$ levels of an alkaline-earth (like) atom is much more significant than the one of the electronic ground states of an ultracold alkali atom. As a result,  to  induce a given  $\Delta_{\rm AC}$, the heating effect of the laser beams for our system is much lower than the ones for the alkali atoms.

This  can be explained as follows.
As mentioned before, here we consider the large-detuning cases where the detuning of the laser is much larger than the fine splitting of the electronic excited states.
For the electronic ground manifold of an alkali atom, all the spin levels are in the same electronic orbit state, i.e., the S-state, and thus have the same dynamical polarizability. Therefore, the spin-dependence of the AC-Stark effect is essentially induced by the  electronic spin-orbit coupling of the excited states \cite{grimm1999optical,zhaihui}.
Thus,
 to realize significant spin-dependence AC-Stark shifts in the  large-detuning cases
one has to apply an extremely strong  beam, and thus the heating effect would be quite large. However, for  an alkaline-earth (like) atom the electronic orbit states corresponding to the $^3$P$_2$ levels $|c,\xi\rangle$  ($\xi=\uparrow,\downarrow,\pm 3/2$) are different from each other.
Precisely speaking, there are  three  electronic orbit P-states that are orthogonal with each other, and the level $|c,\xi\rangle$ corresponds to a $\xi$-dependent probability  mixture of these three orbital states. As a result, these levels have different dynamical polarizability. Therefore, even in the  large detuning cases,  one can still realize very different AC-Stark shifts for these levels
with weak laser beams, and thus the corresponding heating effects can be much weaker.

\begin{figure}
\includegraphics[width=0.35\textwidth]{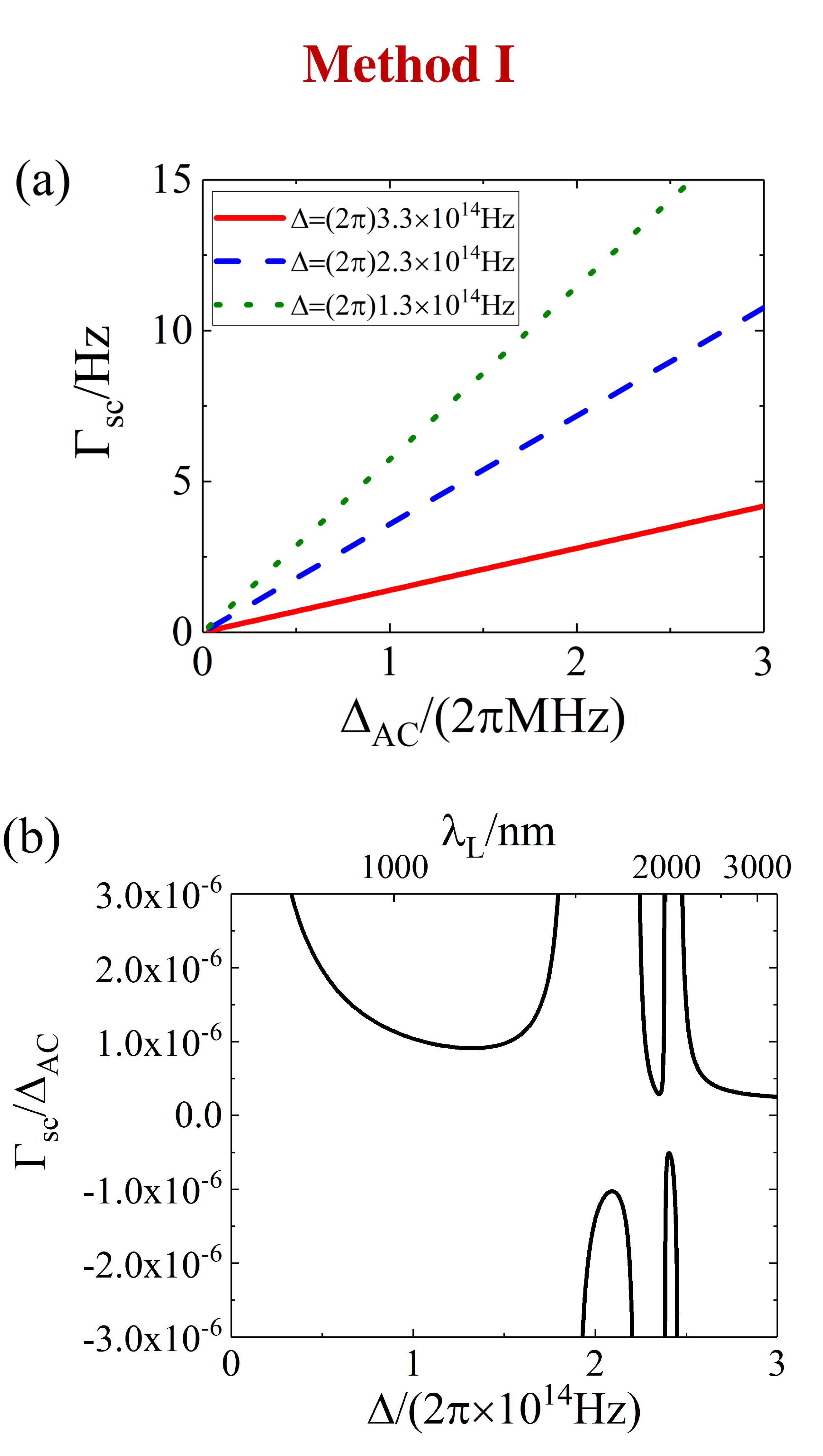}
\caption{(color  online) {\bf (a)} The photon scattering rate $\Gamma_{\rm sc}$ of the system of method I, as a function of the AC-Stark shift difference $\Delta_{\rm AC}$, for the cases with  detuning $\Delta$ of the laser beam with respect to the $^3$P$_2$-$^3$S$_1$ transition (Fig.~\ref{scheme1}(a)) being $\Delta=(2\pi)1.3\times 10^{14}$Hz (green dotted line), $\Delta=(2\pi)2.3\times 10^{14}$Hz (blue dashed line) and  $\Delta=(2\pi)3.3\times 10^{14}$Hz (red solid line).
The laser wavelength corresponding to these cases is $1.15\mu$m, $1.89\mu$m, and $5.08\mu$m, respectively.
 {\bf (b)} The ratio $\Gamma_{\rm sc}/\Delta_{\rm AC}$
as a function of $\Delta$ or the laser wavelength $\lambda_{\rm L}$.}
\label{heating1beam}
\end{figure}

The above discussions yield that for our system one can realize a very large AC-Stark shift difference $\Delta_{\rm AC}$   together with a  long lifetime.
To illustrate this,
we  calculate  $\Delta_{\rm AC}$ and the photon scattering rate $\Gamma_{\rm sc}$ which describes the heating effect, for various cases. The details of this calculation are given in Appendix~\ref{a1}, and the results are shown in Fig.~\ref{heating1beam}.
In the calculation  we take into account the contributions from the excited states $^3$S$_1$ and $^3$D$_{1,2,3}$, which are mostly close to the $^3$P$_2$ levels.
In Fig.~\ref{heating1beam}(a) we illustrate  $\Gamma_{\rm sc}$  as a function of  $\Delta_{\rm AC}$ for the cases where the detuning $\Delta$ of the laser beams with respect to the $^3$P$_2$-$^3$S$_1$ transition (Fig.~\ref{scheme1}(a)) takes various values. It is shown that for
$\Delta=(2\pi)3.3\times 10^{14}$Hz  (corresponding to laser wavelength  5.08$\mu$m)
 we have $\Gamma_{\rm sc}\sim$Hz when $\Delta_{\rm AC}\sim(2\pi)$MHz. If we estimate the lifetime of the ultracold gas as $1/\Gamma_{\rm sc}$, then this result yields that the lifetime of our system can be hundreds of milliseconds. In  Fig.~\ref{heating1beam}(b) we further show the ratio $\Gamma_{\rm sc}/\Delta_{\rm AC}$
as a function of $\Delta$ or the laser wavelength $\lambda_{\rm L}$. Since $\Gamma_{\rm sc}$ is always positive, the sign of this ratio is the same as the one of $\Delta_{\rm AC}$.  In our scheme, $\Delta_{\rm AC}$ is required to be tuned to be positive.
In addition,
the divergence of  $\Gamma_{\rm sc}/\Delta_{\rm AC}$ for $\Delta\approx (2\pi)1.85\times 10^{14}$Hz  is because that we have $\Delta_{\rm AC}=0$ for this particular case. The  divergences of $\Gamma_{\rm sc}/\Delta_{\rm AC}$ for the other four special values of $\Delta$ (including $\Delta=0$) shown in Fig.~\ref{heating1beam}(b)  are due to the resonance between the laser and  the transitions  from $^3$P$_2$ levels to the $^3$S$_1$ and $^3$D$_{1,2,3}$ states for these cases.

\subsection {Effective Inter-Atomic  Interaction}
\label{ss1}

Our scheme is to control the effective SEI of two    atoms in the states $|g,\uparrow\!(\downarrow)\rangle$ and  $|c,\uparrow\!(\downarrow)\rangle$, respectively (Fig.~\ref{scheme1}(b)), with $|g\uparrow\!(\downarrow)$ being defined as the $^1$S$_0$ states with $m_F=-1/2(+1/2)$. In this subsection, we derive the form of the effective interaction between these two atoms.
To this end, we consider the $s$-wave scattering of these two atoms in the zero-energy limit, and perform  discussions in the first-quantization formalism with the two atoms being labeled as 1 and 2.
 In the scattering process these two atoms are incident from the sub-space spanned by the following four  states:
\begin{eqnarray}
|g,\sigma;c,\sigma^\prime\rangle&\equiv&\frac{1}{\sqrt{2}}\left[
|g,\sigma\rangle_1|c,\sigma^\prime\rangle_2-|c,\sigma^\prime\rangle_1|g,\sigma\rangle_2
\right];\nonumber\\
&&\hspace{3.3cm}(\sigma,\sigma^\prime=\uparrow,\downarrow).\label{rgj1}
\end{eqnarray}
Notice that since the atoms are identical fermions, the $s$-wave scattering occurs only when they are in  anti-symmetric internal states. Furthermore, during the scattering process the inter-atomic interaction can couple the states in Eq. (\ref{rgj1}) to the states \cite{4page}

\begin{eqnarray}
|g,{\tilde \sigma};c,\xi\rangle&\equiv&\frac{1}{\sqrt{2}}\left[
|g,{\tilde \sigma}\rangle_1|c,\xi\rangle_2-|c,\xi\rangle_1|g,{\tilde \sigma}\rangle_2
\right],\nonumber\\
&&\hspace{1.5cm}({\tilde \sigma}=\uparrow,\downarrow;\ \ \ \xi=\pm 3/2).\label{rgj2}
\end{eqnarray}
Taking into account  these states, we can express
 Hamiltonian for the two-atom relative motion and internal states  as
\begin{eqnarray}
{\hat H}=-\frac{\nabla^2_{\bf r}}{2\mu}+{\hat H}_f+{\hat V}^{(2)}({\bf r}),\label{h1}
\end{eqnarray}
where $\mu$ and ${\bf r}$ are the reduced mass and relative position of the two atoms, respectively, and ${\hat V}^{(2)}({\bf r})$ is the projection of the interaction potential between one atom in the  $^1$S$_0$ state and another atom in the $^3$P$_{2}(F=3/2)$ states, respectively,
on the subspace spanned by the states in Eqs.~(\ref{rgj1}, \ref{rgj2}).
The explicit form of ${\hat V}^{(2)}({\bf r})$ is given  in Appendix~\ref{int}. Moreover, in Eq.~(\ref{h1}) the operator ${\hat H}_f$ is the free Hamiltonian of the  internal state of the two atoms, which is given by
\begin{eqnarray}
{\hat H}_f&=&\Delta_{\rm AC}\!\!
\sum
_{\tiny
\begin{array}{l}
\sigma=\uparrow,\downarrow\\
\xi=\pm 3/2
\end{array}
}\!\!\!\!
|g,\sigma;c,\xi\rangle\langle g,\sigma;c,\xi|
\label{hc1}
\end{eqnarray}
where the free energy of the states $|g,\sigma;c,\sigma^\prime\rangle$ ($\sigma,\sigma^\prime=\uparrow,\downarrow$) is chosen to be zero.  Here we ignore the electronic $^3$P$_{0,1}$ and $^3$P$_2$ ($F=5/2$) states, because the  energy differences between these states and the ones relevant to our proposal are very large.

In summary, there are four open channels (i.e., the channels corresponding to  $|g,\sigma;c,\sigma^\prime\rangle$ with $\sigma,\sigma^\prime=\uparrow,\downarrow$) and four closed channels (i.e., the channels corresponding to  $|g,\sigma;c,\xi\rangle$ with $\sigma=\uparrow,\downarrow$ and $\xi=\pm3/2$) with the  potential of each channel and
the coupling between different channels
 all being determined by the interaction ${\hat V}^{(2)}({\bf r})$.

Furthermore,
as shown in Appendix~\ref{int}, the interaction potential ${\hat V}^{(2)}({\bf r})$ is anisotropic, and  can couple the wave functions  with the angular momentum of two-atom relative motion being $l$ and $l+2$. Nevertheless, the projection
\begin{eqnarray}
M\equiv m_{F1}+m_{F2}+m_l\label{bigM}
\end{eqnarray}
of the total angular momentum along the $z$-axis is conserved in the scattering process, where $m_{F1,2}$ is the magnetic quantum number of the
atoms 1,2, and $m_l$ is the $z$-component of the angular momentum of  two-atom relative motion. Using this fact and other properties of ${\hat V}^{(2)}({\bf r})$ we find that
 in the zero-energy limit if the two atoms were incident from one of the following four states, i.e., the polarized states $|g,\uparrow;c,\uparrow\rangle$ and $|g,\downarrow;c,\downarrow\rangle$, and the anti-polarized states $|\pm \rangle$ defined by
\begin{eqnarray}
|\pm\rangle\equiv\frac{1}{\sqrt{2}}\big[|g,\uparrow;c,\downarrow\rangle\mp |g,\downarrow;c,\uparrow\rangle\big],\label{pm1}
\end{eqnarray}
then there is only elastic scattering, i.e.,  the two-atom internal state cannot be changed by the scattering process. We denote $|\Psi_{\sigma,\sigma}({\bf r})\rangle$  ($\sigma=\uparrow,\downarrow$) as the zero-energy scattering wave functions corresponding to the polarized incident state $|g,\sigma;c,\sigma\rangle$,
and  $|\Psi_{\pm}({\bf r})\rangle$ as the ones for
 the incident state $|\pm\rangle$ defined in Eq. (\ref{pm1}). The above analysis yields
  \begin{eqnarray}
\lim_{r\rightarrow\infty}|\Psi_{+}({\bf r})\rangle&=&\left(1-\frac{a_{+}}{r}\right)|+\rangle;\label{s11}\\
\lim_{r\rightarrow\infty}|\Psi_{-}({\bf r})\rangle&=&\left(1-\frac{a_{-}}{r}\right)|-\rangle;\\
\lim_{r\rightarrow\infty}|\Psi_{\uparrow,\uparrow}({\bf r})\rangle&=&\left(1-\frac{a_{\rm f}}{r}\right)|g,\uparrow;c,\uparrow\rangle;\\
\lim_{r\rightarrow\infty}|\Psi_{\downarrow,\downarrow}({\bf r})\rangle&=&\left(1-\frac{a_{\rm f}}{r}\right)|g,\downarrow;c,\downarrow\rangle,\label{s41}
\end{eqnarray}
with $a_{\pm}$ and $a_{\rm f}$ being the corresponding elastic scattering lengths. Notice that the scattering lengths for the polarized incident states $|g,\uparrow;c,\uparrow\rangle$ and $|g,\downarrow;c,\downarrow\rangle$ are the same, because our system is invariant under the reflection with respect to the $x-y$ plane.

When the atoms are incident from a superposition of the four special states $|g,\uparrow;c,\uparrow\rangle$, $|g,\downarrow;c,\downarrow\rangle$ and
$|\pm\rangle$, the scattering state would be the corresponding superposition of the ones in Eqs. (\ref{s11}-\ref{s41}), and thus the scattering amplitudes can be expressed in terms of $a_{\pm}$ and $a_{\rm f}$. In particular, when the atoms are incident from the anti-polarized state $|g,\downarrow;c,\uparrow\rangle$ or $|g,\uparrow;c,\downarrow\rangle$, the corresponding  scattering wave function $|\Psi_{\downarrow,\uparrow}({\bf r})\rangle$ or $|\Psi_{\uparrow,\downarrow}({\bf r})\rangle$ satisfy
\begin{eqnarray}
\lim_{r\rightarrow\infty}|\Psi_{\downarrow,\uparrow}({\bf r})\rangle&=&\left[1-\frac{(a_{-}+a_+)/2}{r}\right]|g,\downarrow;c,\uparrow\rangle\nonumber\\
&&-\frac{(a_{-}-a_+)/2}{r}|g,\uparrow;c,\downarrow\rangle,\label{sx11}
\end{eqnarray}
and
\begin{eqnarray}
\lim_{r\rightarrow\infty}|\Psi_{\uparrow,\downarrow}({\bf r})\rangle&=&\left[1-\frac{(a_{-}+a_+)/2}{r}\right]|g,\uparrow;c,\downarrow\rangle\nonumber\\
&&-\frac{(a_{-}-a_+)/2}{r}|g,\downarrow;c,\uparrow\rangle.\label{sx21}
\end{eqnarray}
Eqs. (\ref{sx11}, \ref{sx21}) yield that the spin-exchange scattering process
\begin{eqnarray}
|g,\uparrow;c,\downarrow\rangle\Leftrightarrow |g,\downarrow;c,\uparrow\rangle\label{ie1}
\end{eqnarray}
can occur, and the amplitude of spin-exchange is just $(a_--a_+)/2$.

According to the above discussion, the low-energy scattering between these two atoms can be described by the pseudo potential
\begin{eqnarray}
&&{\hat V}_{\rm eff}=\frac{2\pi}{\mu}\Big[a_{\rm +}|+\rangle\langle+|+a_{-}|-\rangle\langle-|+a_{\rm f}{\hat P}_{\rm f}\Big]\delta({\bf r})\frac{\partial}{\partial r}(r\cdot),\nonumber\\
\label{veff1}
\end{eqnarray}
where
\begin{eqnarray}
{\hat P}_{\rm f}=\sum_{\sigma=\uparrow,\downarrow}|g,\sigma;c,\sigma\rangle\langle g,\sigma;c,\sigma|
\end{eqnarray}
is the projection operator of the polarized states.

In addition, we can treat the electronic states $^1$S$_0$($g$) and $^3$P$_2(F=3/2)$($c$)  as the labels of the two atoms, and treat the $g$- and $c$-atoms as  two distinguishable particles. Furthermore, both of these two atoms are particles with pseudo-spin $1/2$, because in the open channel each atom has two possible magnetic quantum numbers $\uparrow$ and $\downarrow$. In this treatment,
one can express the effective two-atom Hamiltonian in the form  mentioned in Sec.~\ref{introduction}, {\it i.e.},
\begin{eqnarray}
H_{\rm 2body}^{\rm (eff)}=\frac{{\bm p}_S^2}{2m}+\frac{{\bm p}_P^2}{2m}+{\hat V}_{\rm eff}. \nonumber
\end{eqnarray}
Here
${\bm p}_{S(P)}$ is the momentum operator  of the $g$- ($c$-) atom, and
${{\bm p}_S^2}/(2m)+{{\bm p}_P^2}/{(2m)}$ is the {\it pseudo-spin independent free Hamiltonian}, and \begin{eqnarray}
&&{\hat V}_{\rm eff}=\frac{2\pi}{\mu}\times\nonumber\\
&&\left[\!\frac{A_{x}}2\hat{\sigma}_x^{(S)}\hat{\sigma}_x^{(P)}\!
+\!\frac{A_{y}}2\hat{\sigma}_y^{(S)}\hat{\sigma}_y^{(P)}\!
+\!\frac{A_{z}}2\hat{\sigma}_z^{(S)}\hat{\sigma}_z^{(P)}\!+\!A_0
\right]\!
\delta({\bf r})\frac{\partial}{\partial r}(r\cdot)\nonumber\\
\label{mi1}
\end{eqnarray}
is  the effective inter-atomic interaction which is equivalent to the one of Eq~(\ref{veff1}).
Here  ${\hat{\sigma}}_{x,y,z}^{(S(P))}$ are the Pauli operators of the pseudo spin of the $g$-atom ($c$-atom), and the coefficients $A_{x,y,z,0}$ are given by
\begin{eqnarray}
A_{x}&=&A_y\nonumber\\
&=&\frac{a_--a_{+}}{2};\label{aex2}\\
A_z&=&\frac{2a_{\rm f}-(a_-+a_+)}2;\\
A_0&=&\frac{2a_{\rm f}+(a_-+a_+)}4.\label{a02}
\end{eqnarray}

\begin{figure*}
\includegraphics[width=0.7\textwidth]{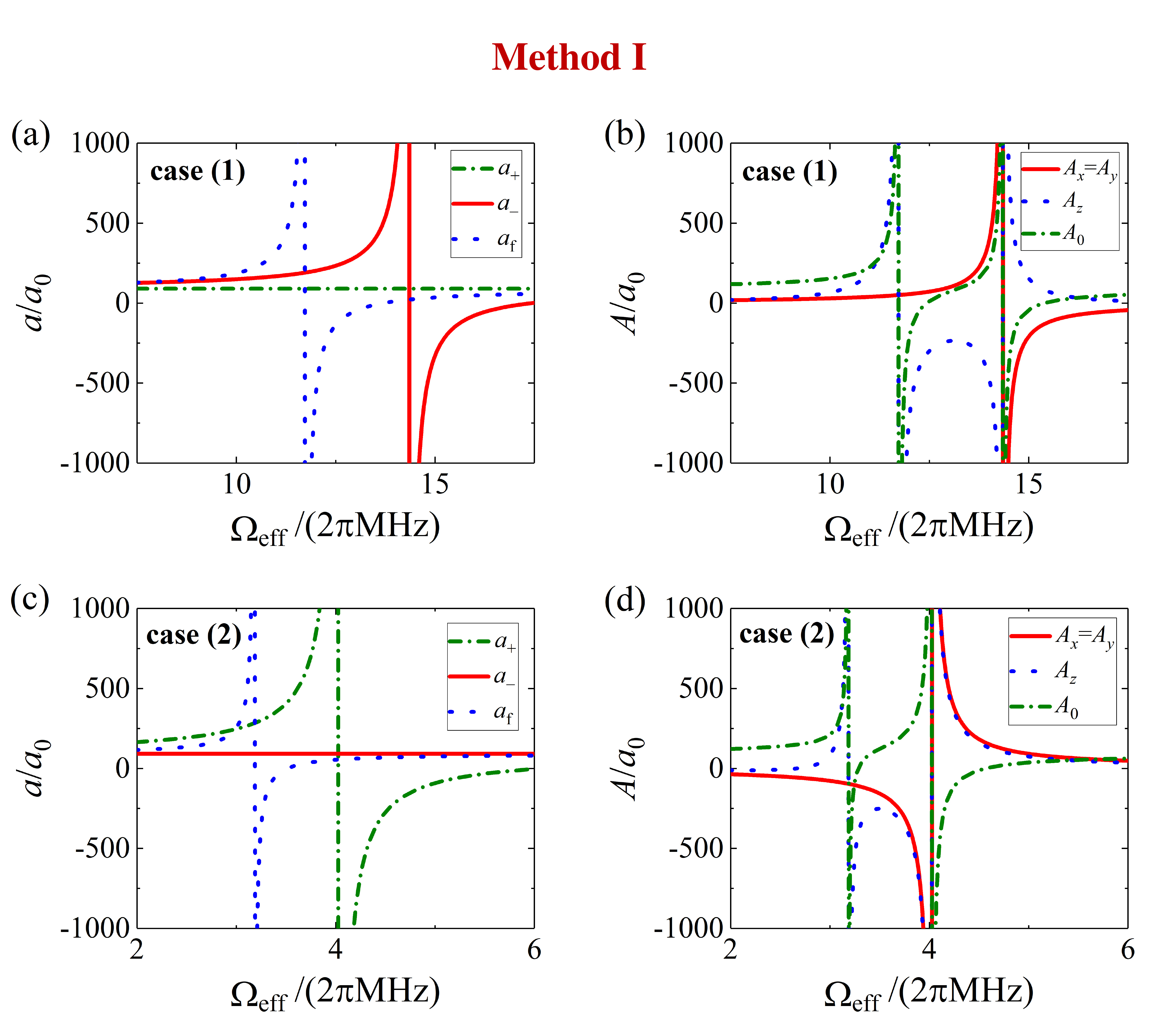}
\caption{(color  online) The scattering lengths $a_{\pm,{\rm f}}$ ({(a)}, {(c)}) and interaction parameters $A_{x,y,z,0}$ ({(b)}, {(d)}) of the system of method I. Here we show the results for two cases, which are given by the multi-channel square-well model  in Sec.~\ref{model1},  with width $b=85a_0$ and other parameters being given in Table I of Appendix~\ref{apppara}.}
\label{sw1}
\end{figure*}

\subsection{Resonant Control of $a_\pm$ and $a_{\rm f}$}
\label{2c}

Now we show that the  scattering lengths $a_\pm$ and $a_{\rm f}$ can be resonantly controlled by the AC-Stark shift difference $\Delta_{\rm AC}$.

As mentioned above, the open channels corresponding to  $|g,\sigma;c,\sigma^\prime\rangle$ ($\sigma,\sigma^\prime=\uparrow,\downarrow$) is coupled to the closed channels corresponding to  $|g,\sigma;c,\xi\rangle$ ($\sigma=\uparrow,\downarrow$ and $\xi=\pm 3/2$) by the anisotropic interaction ${\hat V}^{(2)}({\bf r})$.
Explicitly, due to the conservation of the angular momentum $M$ defined in Eq.~(\ref{bigM}), the $s$-wave states of the open channels of our system are coupled to the  $d$-wave states of the closed channels.
Therefore, the scattering lengths $a_{\pm,{\rm f}}$ depend on the energy gaps between the thresholds of the open and closed channels, which is just the AC-Stark shift $\Delta_{\rm AC}$ (Fig.~\ref{scheme1}(c)), and thus one can control these scattering lengths through $\Delta_{\rm AC}$ by changing the intensities of the  laser beam.

Furthermore, as shown before, the order of magnitude of   $\Delta_{\rm AC}$ can be as large as $(2\pi)$MHz, with the photon scattering rate $\Gamma_{\rm sc}$ being only of the order of Hz.
On the other hand, the
characteristic energy corresponding to the length scale of the inter-atomic van der Waals interaction potential, i.e., the  van der Waals energy $E_{\rm vdW}$, is also of this order for  Yb atoms \cite{one6page}.
Therefore, if the closed channels have $d$-wave bound states with
binding energy $|E_\uparrow|$ comparable or less than $E_{\rm vdW}$, by tuning $\Delta_{\rm AC}$ via the laser intensity  one can make
the open channels to be near-resonant to the closed-channel bound states, i.e., realize Feshbach resonances, while keeping the heating effect being low enough.
At each resonance point, one of  the three scattering lengths $a_{\pm,{\rm f}}$ diverges.
Around the resonances, one can efficiently manipulate
 $a_{\pm,{\rm f}}$ (or the interaction parameters $A_{x,y,z,0}$). That is the basic principle of this method.

\subsection{Illustration with Multi-Channel Square-Well Model}
\label{model1}

We illustrate our approach with a calculation for the scattering lengths $a_{\pm,{\rm f}}$ and the interaction parameters $A_{x,y,z,0}$ defined above.
As shown in Appendix~\ref{int}, the potential ${\hat V}^{(2)}({\bf r})$ can be expressed as
\begin{eqnarray}
{\hat V}^{(2)}({\bf r})=\sum_{\eta=1}^6V^{(2)}_{\eta}(r){\hat D}_{\eta}({\hat{\bf r}}),
\end{eqnarray}
 where $r=|{\bf r}|$ and ${\hat{\bf r}}={\bf r}/r$. Here $V^{(2)}_{1,...,6}(r)$ are the potential curves corresponding to six different electronic states. These electronic states are defined in Appendix~\ref{int}, together with the
operators ${\hat D}_{1,...,6}({\hat{\bf r}})$. As mentioned before, we do not know the parameters of the potential curves $V^{(2)}_{1,...,6}(r)$. Therefore,  we can only qualitatively illustrate our proposal with a
multi-channel square-well model with
\begin{eqnarray}
V^{(2)}_{\eta}(r)=-U^{(2)}_{\eta}\theta(b-r),\ \ \  (r\geq 0;\ \eta=1,...,6),
\label{v2sw}
\end{eqnarray}
where  $\theta(x)$ is the step function satisfying $\theta(x)=1$ for $x\geq 0$ and $\theta(x)=0$ for $x< 0$. In our calculation, we choose $b=85a_0$, with $a_0$ being the Bohr radius and  $2b$ taking a typical value of the length scale $\beta_6$ associated with the van der Waals interaction between atoms as heavy as Ytterbium \cite{two6page}.
We consider all the involved $s$-wave and $d$-wave channels, and ignore the channels with higher relative angular momentum, for simplicity. We also
ignore the centrifugal potential of the $d$-wave channels in the region $r<d$ because the square-well potentials in this region are very deep.

We display the results for two cases in Fig.~\ref{sw1}. In  Table I of Appendix~\ref{apppara} we show the
value of the potential depth $U^{(2)}_{\eta}$ $(\eta=1,...,6)$ as well as the $s$-wave scattering length $a_{\eta}^{(2)}$ corresponding to a single-channel square-well potential $V^{(2)}_{\eta}(r)$ given in Eq.~(\ref{v2sw}).
In Fig.~\ref{sw1} we illustrate  the behaviors of both the scattering lengths $a_{\pm,{\rm f}}$ and the interaction parameters $A_{x,y,z,0}$ defined in Eqs.~(\ref{aex2}-\ref{a02})  as functions of $\Delta_{\rm AC}$. It is shown that  using the resonances one can tune the intensity $A_{x}=A_y$ of the effective SEI as well
as the intensities $A_{z,0}$ of the spin-non-exchanging and spin-independent interaction in a broad region {\it e.g.}, between $-1000a_0$ and $1000a_0$ through the laser intensity, and may prepare the effective inter-atomic interaction ${\hat V}_{\rm eff}$ to be either ``anti-ferromagnetic like" ($A_{x}=A_y>{\rm Max}[0,-A_z]$) with the lowest eigen state being the singlet state $|+\rangle$ defined in Eq. (\ref{pm1}), or ``ferromagnetic like" ($A_z<0$ and $|A_{x}|=|A_y|<|A_z|$) with the  lowest eigen states being the polarized ones $|g,\uparrow;c,\uparrow\rangle$ and $|g,\downarrow;c,\downarrow\rangle$. It is also possible to ``turn off" the spin-independent interaction by tuning $A_0=0$ or prepare the system to other required interaction parameter regions.
In Appendix~\ref{appd} we show the values of $\Delta_{\rm AC}$ under which we have  $A_0=0$ for the  two cases in Fig.~\ref{sw1}, as well as the corresponding values of  $A_{x,y,z}$ and the scattering lengths $a_{\pm, f}$.

\section{Method II}
\label{nm2}

\begin{figure}
\includegraphics[width=0.46\textwidth]{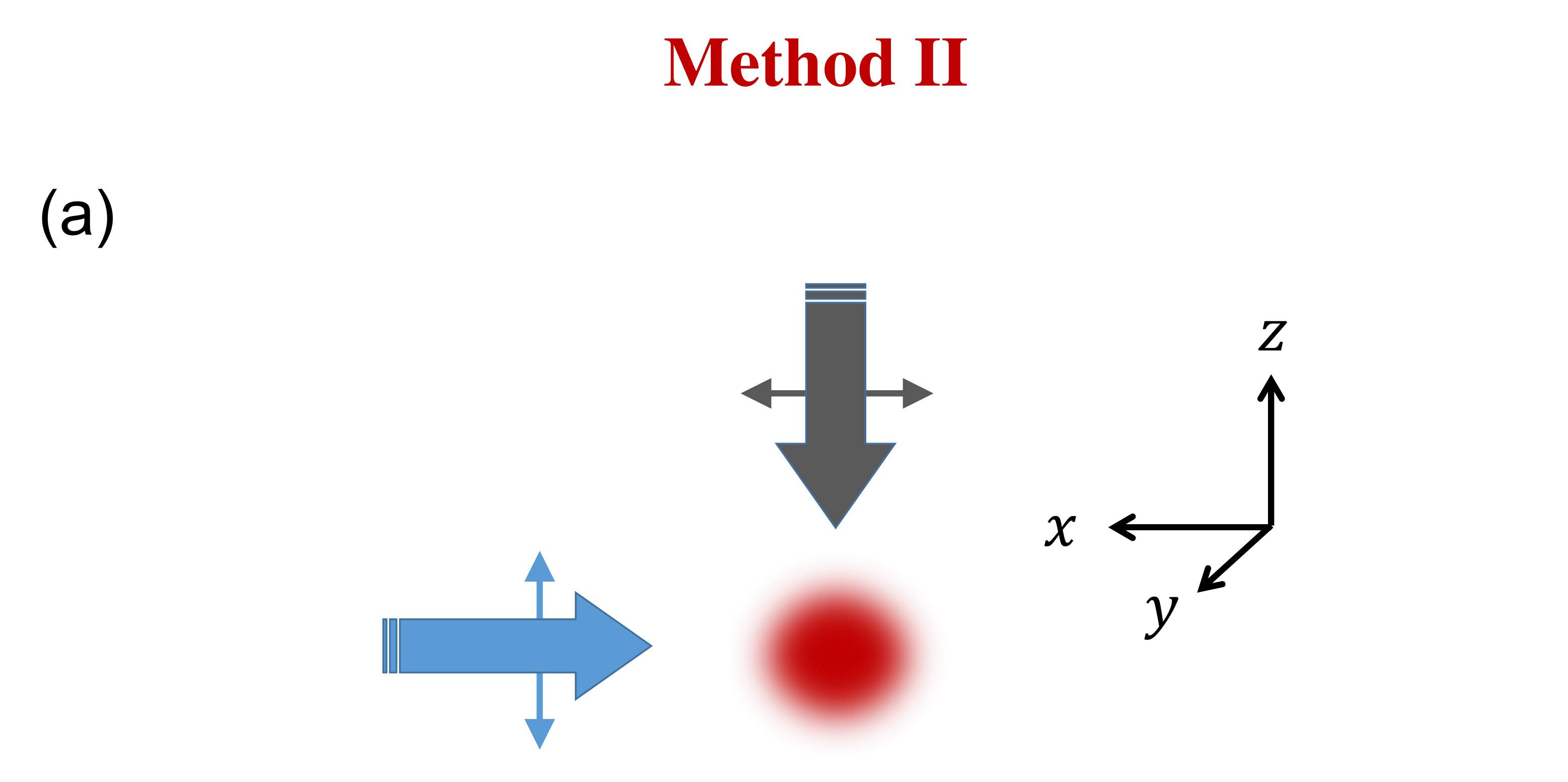}\\
\vspace{0.5cm}

\includegraphics[width=0.46\textwidth]{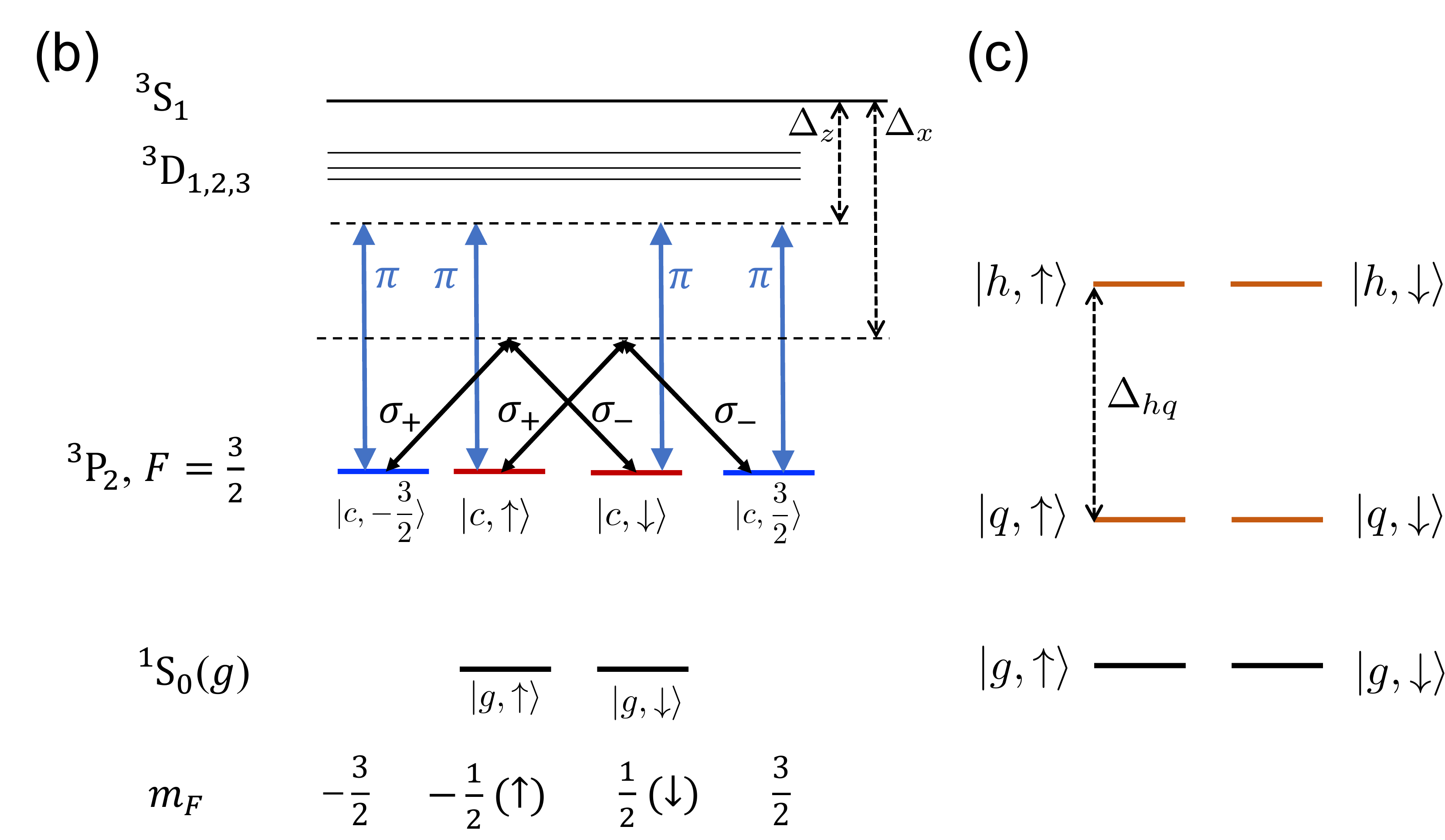}\\ 
\vspace{0.5cm}

\includegraphics[width=0.46\textwidth]{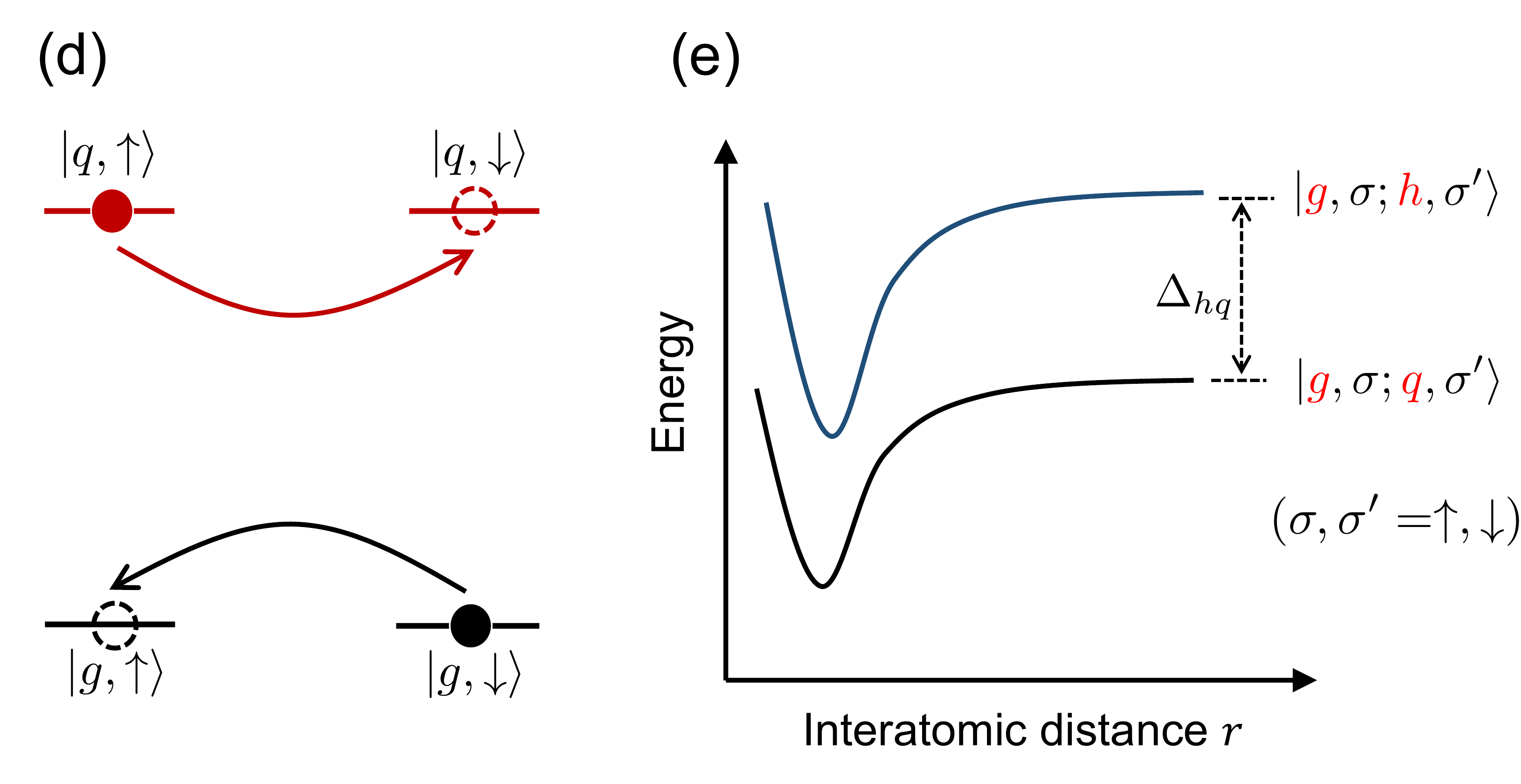} 
\caption{(color  online) Schematic  illustration for method II. {\bf  (a):}
Two laser beams polarized along the $z$-direction and the $x$-direction, respectively, are applied to the atoms. {\bf  (b):}
1-atom bare levels as well as the influences of the laser beams. Here the beam polarized along the $x$-direction is decomposed into two  beams with $\sigma_+$ and $\sigma_-$ polarizations, respectively.  {\bf  (c):} 1-atom dressed levels in the rotating frame. {\bf  (d):} A 2-atom spin-exchange process. The black (red) filled and dashed circles represent the $g(q)-$atom before and after a collision, respectively. Both this process and the inverse one are studied in this work. {\bf  (e):}The  inter-atomic scattering channels.
 Here the solid curves represent the potentials of each channel. The
 coupling potentials between different channels are not shown in the figure.}
\label{schemeimprove}
\end{figure}

\begin{figure}
\includegraphics[width=0.38\textwidth]{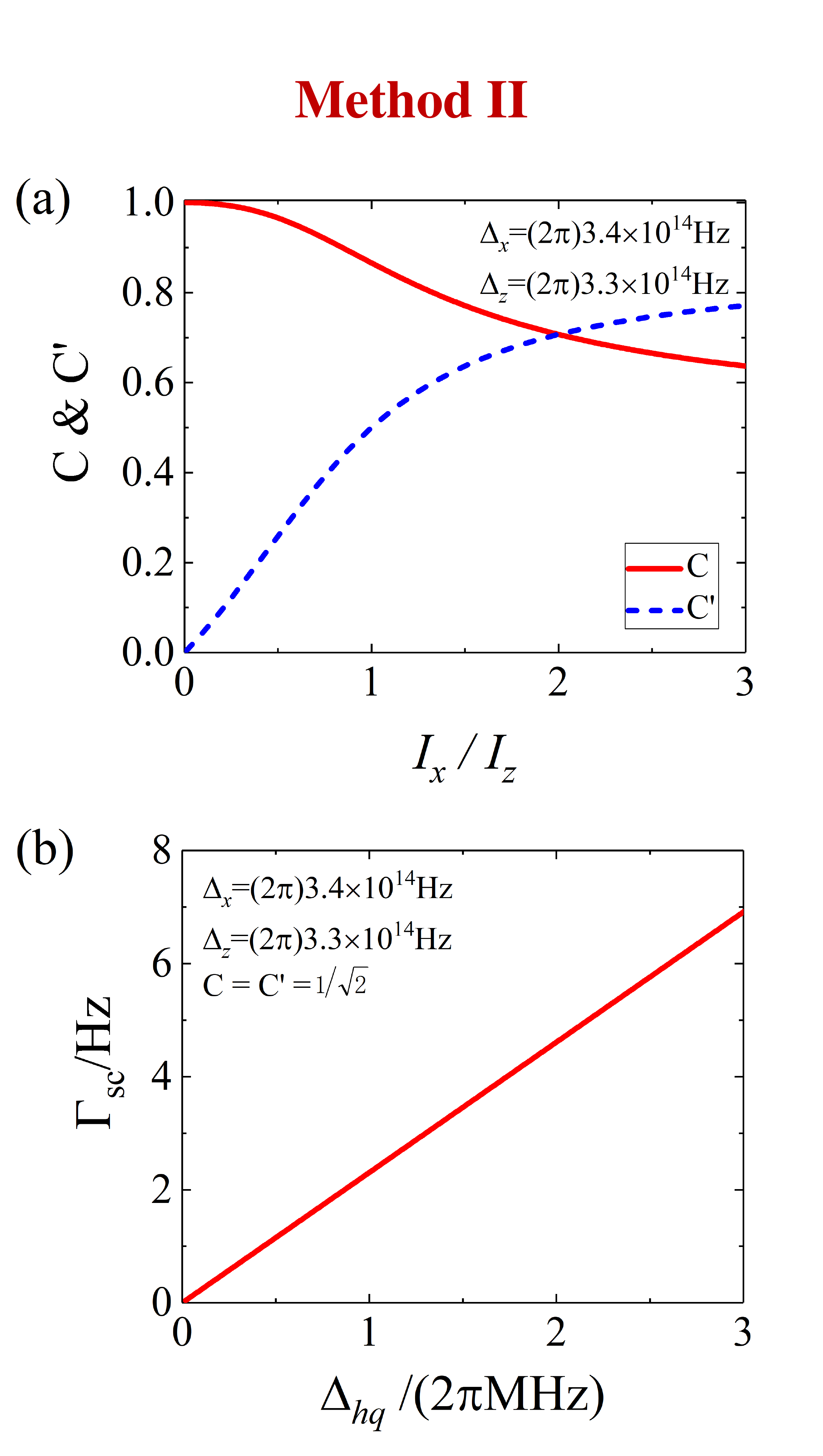}
\caption{(color  online) {\bf(a):} The parameters $C$ and $C^\prime$ in Eqs.~(\ref{hq3}-\ref{hq2})
as a function of $I_x/I_z$, with $I_{z(x)}$ being the intensity of the laser beam polarized along the $z$-($x$-)direction. Here we show the results
for the cases
with  $\Delta_{z}=(2\pi)3.3\times 10^{14}$Hz  and $\Delta_{x}=(2\pi)3.4\times 10^{14}$Hz,
 where $\Delta_{z(x)}$ is the  detuning of the $z(x)$-polarized  laser beam with respect to the $^3$P$_2$-$^3$S$_1$ transition (Fig.~\ref{schemeimprove}(a)).
 In this case the  wave length of the  $z(x)$-polarized  laser beam  is  $5.08\mu$m ($6.12\mu$m).
 {\bf(b):} The total photon scattering rate $\Gamma_{\rm sc}$ as a function of the energy gap  $\Delta_{hq}$  between the higher  dressed states $|h,\uparrow\!(\downarrow)\rangle$ and  the lower dressed states $|q,\uparrow\!(\downarrow)\rangle$. Here we show the results for the cases with $C^\prime=C=1/\sqrt{2}$, and the frequencies of the   laser beams polarized along the $z$- ($x$-) direction being the same as (a).
}
\label{hh2}
\end{figure}

Now we introduce our second approach for the manipulation of SEI.
As above, we also take the system of two $^{171}$Yb atoms as an example.  In addition, to avoid using too many different symbols, in this  and the next section, we will use some notations which have already been used in Sec.~\ref{m1}, for the clear-enough cases   ({\it e.g.,} we still use $\Gamma_{\rm sc}$ to denote the photon scattering rate of laser beams for the current scheme). The exact definitions of these notations  for this section will all be given in the following discussions.

\begin{figure*}
\includegraphics[width=0.7\textwidth]{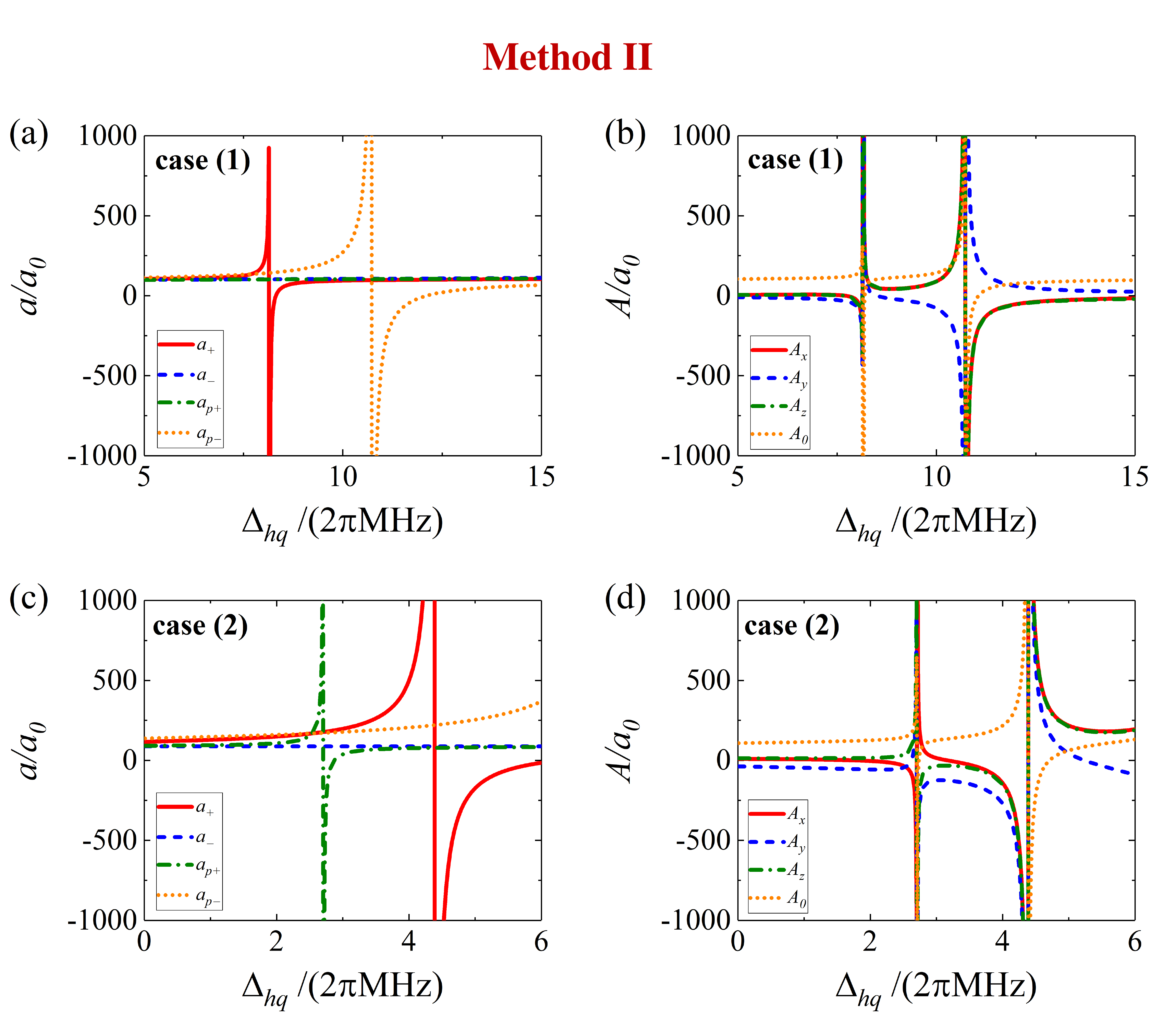}
\caption{(color  online) The scattering lengths $a_{\pm}$ and $p_{\pm}$ ({(a)}, {(c)}) and interaction parameters $A_{x,y,z,0}$ ({(b)}, {(d)}) of the system of method II. Here we show the results for two cases, which are given by the multi-channel square-well model  in Sec.~\ref{model1},  with width $b=85a_0$ and other parameters being given in Table II of Appendix~\ref{apppara}. The values of $\Delta_{hq}$ under which we have  $A_0=0$, as well as the corresponding values of  $A_{x,y,z}$ and $a_{\pm, p\pm}$, are shown in Appendix~\ref{appd}.}
\label{rsw2}
\end{figure*}

%\subsection{System and Setup}

Method II is a generalization of method I. In this approach
two laser beams are applied, which are polarized along the $z$-direction and the the $x$-direction, respectively (Fig.~\ref{schemeimprove}(a)).
Both of these two beams are far-off resonant for the transition from the  $^3$P$_2(F=3/2)$  level to the excited states ({\it i.e.},  the detunings  are much larger than the  Rabi frequencies of the transitions and the natural linewidth of the excited states), similar as in Sec.~\ref{m1}. The frequency difference of these two beams is not
 required to take any certain value.   The only requirement for this frequency difference is that it
 should be much larger than the AC-Stark shifts induced by each beam so that the two beams induce the second-order effects for the  $^3$P$_2(F=3/2)$ level independently. In this case, the total effect of these two beams is {\it not} equivalent to the one of a single beam polarized along some direction between the $x$- and $z$-axis.

As in Sec.~\ref{m1a},
the  beam polarized along the $z$-direction can induce spin-dependent AC-Stark shifts for the states in the $^3$P$_2(F=3/2)$ level. Meanwhile, the beam polarized along the $x$-direction can be decomposed into two  beams with $\sigma_+$ and $\sigma_-$ polarizations, respectively, and thus
induces Raman coupling between the $^3$P$_2(F=3/2)$ state with $m_F=-3/2(+3/2)$ and the one with $m_F=1/2(-1/2)$ (Fig.~\ref{schemeimprove}(b)). As a result, four dressed states $|q(h),\uparrow\!(\downarrow)\rangle$ can be formed (Fig.~\ref{schemeimprove}(c)), which can be expressed as
\begin{eqnarray}
|h,\uparrow\rangle&=&C^\prime|c,\uparrow\rangle+C|c,3/2\rangle, \label{hq3}\\
|q,\uparrow\rangle&=&-C|c,\uparrow\rangle+C^\prime|c,3/2\rangle;\label{hq1}
\end{eqnarray}
and
\begin{eqnarray}
|h,\downarrow\rangle&=&C^\prime|c,\downarrow\rangle+C|c,-3/2\rangle,\label{hq4}\\
|q,\downarrow\rangle&=&-C|c,\downarrow\rangle+C^\prime|c,-3/2\rangle;\label{hq2}
\end{eqnarray}
where  $|c,\sigma\rangle$ ($\sigma=\uparrow,\downarrow,\pm 3/2$) are just the $^3$P$_2(F=3/2)$ states, as defined in Sec.~\ref{m1} and shown in (Fig.~\ref{schemeimprove}(b)).  The
coefficients $C$ and $C^\prime$ in Eqs.~(\ref{hq3}-\ref{hq2}),
 are determined by the intensities and frequencies of the two laser beams, with the expressions being derived in Appendix~\ref{nmm2}.  In Fig.~\ref{hh2}(a) we show the values of $C$ and $C^\prime$ as functions of the  ratio between the intensities of the two laser beams for typical cases. Furthermore, due to the reflection symmetry with respect to the $x-y$ plane, the two lower (higher) dressed states $|q(h),\uparrow\rangle$ and  $|q(h),\downarrow\rangle$ are degenerate (Fig.~\ref{schemeimprove}(c)).
The
 energy gap $\Delta_{hq}$   between these higher and lower dressed states (Fig.~\ref{schemeimprove}(c)) is also derived in
Appendix~\ref{nmm2}, where we find that $\Delta_{hq}$ can be expressed as
\begin{eqnarray}
\Delta_{hq}=\sqrt{\left[\Delta_{\rm AC}^{(z)}-\Delta_{\rm AC}^{(x)}/2\right]^2+\frac34\Delta_{\rm AC}^{(x)2}},\label{dhqq}
\end{eqnarray}
where $\Delta_{\rm AC}^{(z(x))}$ is the AC-Stark shift difference only induced by the
 laser beams polarized along the $z$- ($x$-) direction, as defined in Sec.~\ref{m1}. Therefore, $\Delta_{hq}$ has the same order of magnitude with  the AC-Stark shift difference $\Delta_{\rm AC}^{(z(x))}$, as mentioned in Sec.~\ref{introduction}.

Furthermore, the total photon scattering rate $\Gamma_{\rm sc}$ of these two  beams, which describes the heating effects, can also be calculated directly (Appendix~\ref{nmm2}).
In Fig.~\ref{hh2}(b) we show $\Gamma_{\rm sc}$ as a function of $\Delta_{hq}$ for a typical case. It is shown that we have $\Gamma_{\rm sc}\sim$Hz when $\Delta_{hq}\sim(2\pi)$MHz. Therefore, the laser-induced heating effect is weak, which is similar to the one in Sec.~\ref{m1}.

% For instance, when the frequency difference of these two beams is much less than the detuning of these beams with respect to the transitions from the  $^3$P$_2(F=3/2)$  level to the excited states, and the intensity of the $\pi$ beam polarized along the $z$ is half of the one polarized along the $x$-axis, we have  $C_q=-C_q^\prime=1/\sqrt{2}$ and $C_h=C_h^\prime=1/\sqrt{2}$, and the total photon scattering rate $\Gamma_{\rm sc}$ of the two laser beams, which is the summation of the one of each beam, is $\Gamma_{\rm sc}\approx 3.8\times 10^{-7}\Delta_{hq}$ (i.e., $\Gamma_{\rm sc}\approx 2.4$Hz for $\Delta_{hq}=(2\pi)10^6$Hz).

%\subsection{Effective Interaction }

Furthermore, similar as in Sec.~\ref{m1}, our method is to control the SEI between two atoms in the $^1$S$_{0}$ state ($g$-state) and the lower dressed state ($q$-state) (Fig.~\ref{schemeimprove}(d)). To this end, we consider the scattering  processes in the zero-energy limit, with  incident states being in the Hilbert space spanned by the   states:
\begin{eqnarray}
|g,\sigma;q,\sigma^\prime\rangle&\equiv&\frac{1}{\sqrt{2}}\left[
|g,\sigma\rangle_1|q,\sigma^\prime\rangle_2-|q,\sigma^\prime\rangle_1|g,\sigma\rangle_2
\right];\nonumber\\
&&\hspace{3.3cm}(\sigma,\sigma^\prime=\uparrow,\downarrow).\label{gj1}
\end{eqnarray}
 In addition, the inter-atomic interaction $V^{(2)}({\bf r})$ couples these open channels corresponding to the above four states
 to the closed channels corresponding to
 \begin{eqnarray}
|g,\sigma;h,\sigma^\prime\rangle&\equiv&\frac{1}{\sqrt{2}}\left[
|g,\sigma\rangle_1|h,\sigma^\prime\rangle_2-|h,\sigma^\prime\rangle_1|g,\sigma\rangle_2
\right],\nonumber\\
&&\hspace{3.3cm}(\sigma,\sigma^\prime=\uparrow,\downarrow;).
\end{eqnarray}
The energy gap between the above open and closed channels
 is just $\Delta_{hp}$, as shown in Fig.~\ref{schemeimprove}(e).

In addition, using an analysis based on the properties of the interaction potential ${\hat V}^{(2)}({\bf r})$, which is similar to the discussion in Sec.~\ref{ss1}, we find that in the zero-energy limit if the two atoms were incident from one of the following four states:
\begin{eqnarray}
|\pm\rangle&\equiv&\frac{1}{\sqrt{2}}\big[|g,\uparrow;q,\downarrow\rangle\mp |g,\downarrow;q,\uparrow\rangle\big];\label{ppm1}\\
|p_{\pm}\rangle&\equiv&\frac{1}{\sqrt{2}}\big[|g,\uparrow;q,\uparrow\rangle\mp |g,\downarrow;q,\downarrow\rangle\big],\label{ppm2}
\end{eqnarray}
then in the zero-energy limit there are only elastic scattering processes in which the two-atom internal state is not changed. We denote the elastic scattering lengths with respect to the incident states $|\pm\rangle$ and $|p_{\pm}\rangle$ as $a_{\pm}$ and $a_{ p \pm}$, respectively. Notice that for the current system it is possible that $a_{ p +}\neq a_{ p -}$, i.e., the spin-change processes
$|g,\uparrow,q;\uparrow\rangle\Leftrightarrow|g,\downarrow;q,\downarrow\rangle$ are also permitted.
Therefore, the low-energy interaction between these two atoms can be described
by the pseudo potential
\begin{eqnarray}
{\hat V}_{\rm eff}=\frac{2\pi}{\mu}&&\Big[a_{\rm +}|+\rangle\langle+|+a_{-}|-\rangle\langle-|\nonumber\\
&&+a_{p+}|p_+\rangle\langle p_+|+a_{p-}|p_-\rangle\langle p_-|\Big]\delta({\bf r})\frac{\partial}{\partial r}(r\cdot).\nonumber\\
\label{veff2}
\end{eqnarray}
Moreover, by treating the atoms in $|g,\uparrow\!(\downarrow)\rangle$ and  $|q,\uparrow\!(\downarrow)\rangle$ as  two distinguished particles with pseudo-spin $1/2$,
one can express the effective two-atom Hamiltonian in the form of Sec.~I, {\it i.e.},
\begin{eqnarray}
H_{\rm 2body}^{\rm (eff)}=\frac{{\bm p}_S^2}{2m}+\frac{{\bm p}_P^2}{2m}+{\hat V}_{\rm eff}. \nonumber
\end{eqnarray}
Here
${\bm p}_{S(P)}$ is the momentum operator  of the $g$- ($q$-) atom and
${{\bm p}_S^2}/(2m)+{{\bm p}_P^2}/{(2m)}$ is the {\it pseudo-spin independent free Hamiltonian}, and the effective inter-atomic interaction
\begin{eqnarray}
&&{\hat V}_{\rm eff}=\frac{2\pi}{\mu}\times\nonumber\\
&&\left[\!\frac{A_{x}}2\hat{\sigma}_x^{(S)}\hat{\sigma}_x^{(P)}\!
+\!\frac{A_{y}}2\hat{\sigma}_y^{(S)}\hat{\sigma}_y^{(P)}\!
+\!\frac{A_{z}}2\hat{\sigma}_z^{(S)}\hat{\sigma}_z^{(P)}\!+\!A_0
\right]\!
\delta({\bf r})\frac{\partial}{\partial r}(r\cdot) \nonumber\\
\label{mig1}
\end{eqnarray}
is equivalent to the one of  Eq.~(\ref{veff2}), with ${\hat{\sigma}}_{x,y,z}^{(S(P))}$ being the Pauli operators of the pseudo spin of the $g$- ($q$-) atom.
In Eq.~(\ref{mig1}) the coefficients $A_{x,y,z,0}$ are related to the scattering lengths $a_{\pm}$ and $a_{ p \pm}$ via
\begin{eqnarray}
A_{x}&=&\frac{(a_--a_{+})+(a_{p-}-a_{p+})}{2};\\
A_{y}&=&\frac{(a_--a_{+})-(a_{p-}-a_{p+})}{2};\\
A_z&=&\frac{(a_{p-}+a_{p+})-(a_-+a_+)}2;\\
A_0&=&\frac{(a_{p-}+a_{p+})+(a_-+a_+)}4.
\end{eqnarray}
Notice that
 in the current system the interaction parameters $A_x$ and $A_y$ may be unequal.
%This effective interaction has one more term than the ones in Eqs.~(\ref{mi1}) and (\ref{aeff2}),  i.e., the term proportional to $A_{\rm ex}^\prime$, which describes the coupling between $|g,\uparrow;q,\uparrow\rangle$ and $|g,\downarrow;q,\downarrow\rangle$. Here we emphasis that although this term makes the system more complicated, the total peseudo-spin of the two atoms is still conserved for the current effective interaction, which yields that the triplet and singlet states are note coupled with each other. In addition, ${\hat V}_{\rm eff}$ is still invariant under the simultaneous flipping of the peseudo-spin of the two atoms.

 Similar to before, by changing
 the intensities of the two laser beams, one can tune the energy gap $\Delta_{hp}$ and  induce Feshbach resonances.
The scattering lengths $a_{\pm}$ and $a_{p \pm}$ or the interaction parameters
 $A_{x,y,z,0}$  can be efficiently manipulated via these Feshbach resonances. We illustrate these resonances via the multi-channel square-well model used in Sec.~\ref{model1},
  with width $b=85a_0$ and other parameters being given in Table II of Appendix~\ref{apppara}. The results are shown in Fig.~\ref{rsw2}.

We emphasis that there is an important difference between the current approach and method I. For the system of method I, the $s$-wave states of the open channels are  coupled only to the $d$-wave  states of the closed channels, as mentioned in the above sections. However, for our current system, the $s$-wave open-channel states are coupled to both the $d$-wave and the $s$-wave states of the closed channels. As discussed in Sec.~\ref{introduction}, this fact implies that the  Feshbach resonances for $a_{\pm}$ with a low-enough heating rate are much more possible  to appear for realistic systems. That is an important advantage of the current method.

\section{ Method III}
\label{m2}

Now we introduce our third approach for the manipulation of SEI, which is based on the $^3$P$_2$-$^3$P$_0$
Raman coupling, by  taking $^{171}$Yb atoms as an example. As before, we will use some notations which have  been used in Sec.~\ref{m1}, to reduce the number of different symbols.
The exact definitions of these notations  for this section will be given below.

\subsection{Low-Heating Raman Coupling between $^3$P$_0$ and $^3$P$_2$ Levels}
\label{1b}

\begin{figure}
\includegraphics[width=0.49\textwidth]{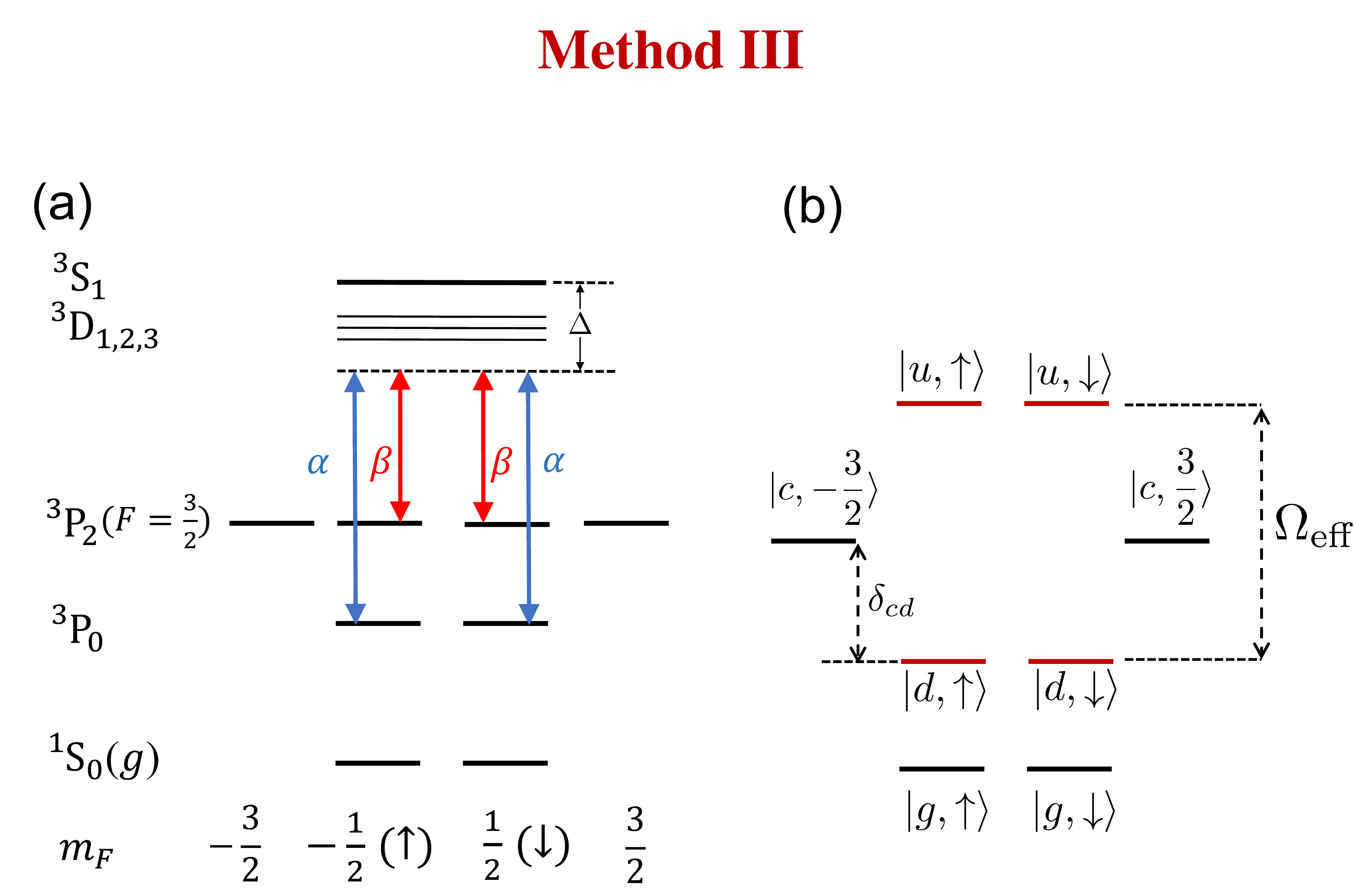}\\
\vspace{1cm}
\includegraphics[width=0.49\textwidth]{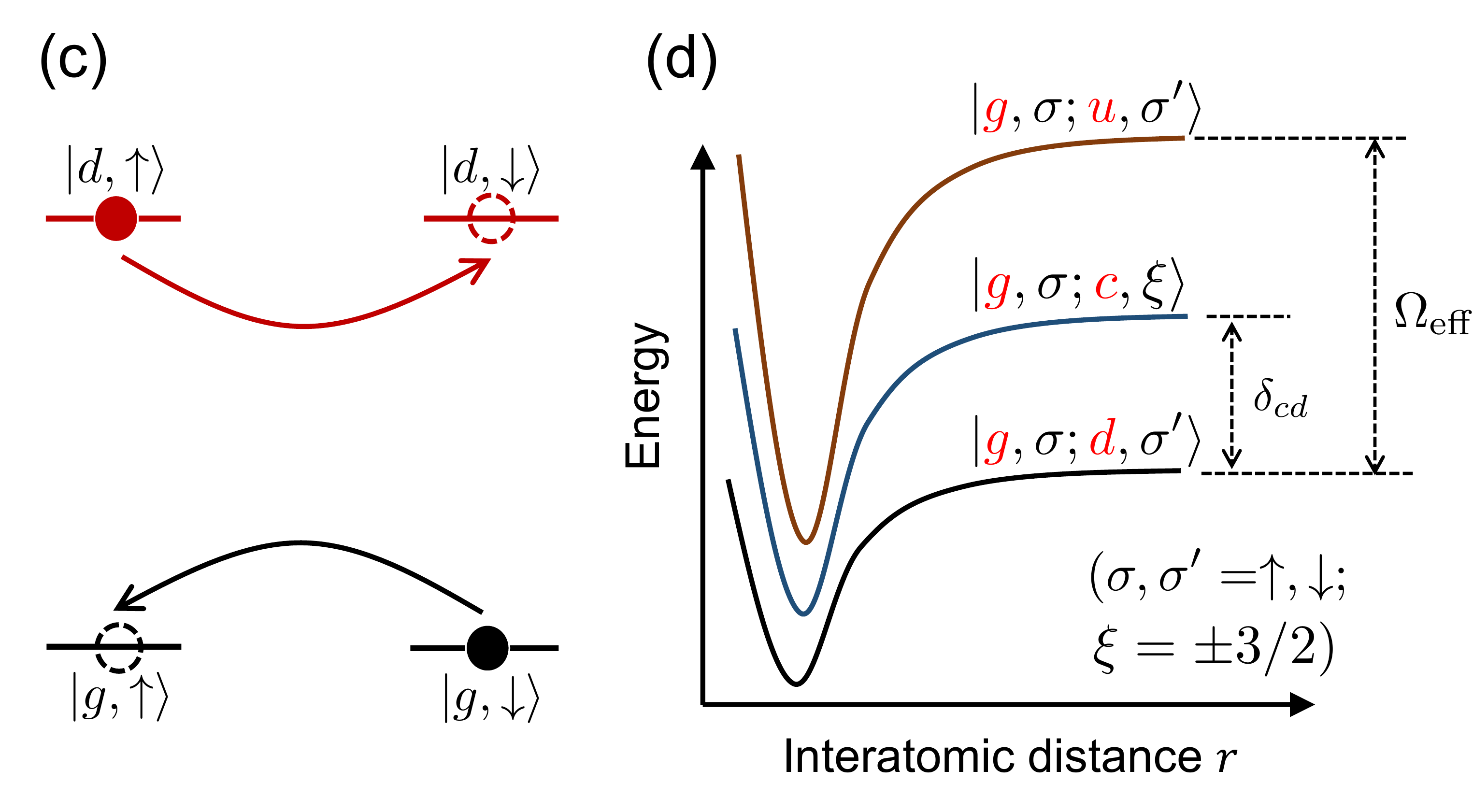} 
\caption{(color  online) Schematic  illustration of method III. {\bf  (a):} 1-atom bare levels and $\pi$-polarized Raman beams (blue and red lines). Here $\Delta$ is the single-photon detuning. {\bf  (b):} 1-atom dressed levels in the rotating frame. {\bf  (c):} A 2-atom spin-exchange process. The black (red) filled and dashed circles represent the $g(d)-$atom before and after a collision, respectively. Both this process and the inverse one are studied in this work. {\bf  (d):}The  inter-atomic scattering channels.
 Here the solid curves represent the potentials of each channel. The
 coupling potentials between different channels are not shown in the figure.}
\label{scheme}
\end{figure}

As shown in Fig.~\ref{scheme}(a), in the current method  two $\pi$-polarized laser beams $\alpha$ and $\beta$ are applied  at a zero magnetic field
so that the $^3$P$_{0,2}$ levels
are far-off resonantly coupled
 to the excited states.
 These two beams can induce
a  Raman coupling between the   states $|^3{\rm P}_0,1/2,\sigma\rangle$ and  $|^3{\rm P}_2,3/2,\sigma\rangle$, with  $\sigma=-1/2(\uparrow)$ or $1/2(\downarrow)$.
%As shown below, the effective Rabi frequency $\Omega_{\rm eff}$ of this Raman coupling can be as large as $(2\pi)$MHz.
Furthermore, the frequency difference of these two beams is tuned to compensate for the difference of the AC-Stark shifts of these  states, so that this Raman coupling is  resonant. Explicitly,  the fluctuation of this frequency difference should be much less than the effective Rabi frequency $\Omega_{\rm eff}$ of the Raman coupling, which is of the order of $(2\pi)$MHz as shown below.

As a result of this resonant Raman coupling,
%in the rotating frame the single-atom Hamiltonian $H_{\rm 1b}$ is given by (Appendix A, $\hbar=1$):
%\begin{eqnarray}
%H_{\rm 1b}&=&E_{^1{\rm S}_0}{\cal P}_{^1{\rm S}_0}+E_{^3{\rm P}_2}\left[{\cal P}_{^3{\rm P}_2}+{\cal P}_{^3{\rm P}_0}\right]\nonumber\\
%&&+\frac{\Omega_{\rm eff}}{2}\sum_{\sigma=\uparrow,\downarrow}\left[|^3{\rm P}_0,1/2,\sigma\rangle|^3{\rm P}_2,3/2,\sigma|+h.c.\right],\nonumber\\
%\end{eqnarray}
%where $E_{^1{\rm S}_0}$ ($E_{^3{\rm P}_2}$) is the energy of $^1{\rm S}_0$ ($^3{\rm P}_2$) levels in the absence of Raman beams,
the eigen states of the single-atom Hamiltonian $H_{\rm 1b}$ in the rotating frame are given by (Appendix~\ref{mm2}):
\begin{eqnarray}
|d,\sigma\rangle&\equiv& \frac{1}{\sqrt{2}}\left[|^3{\rm P}_0,1/2,\sigma\rangle-|^3{\rm P}_2,3/2,\sigma\rangle\right];\label{sd}\\
|u,\sigma\rangle&\equiv& \frac{1}{\sqrt{2}}\left[|^3{\rm P}_0,1/2,\sigma\rangle+|^3{\rm P}_2,3/2,\sigma\rangle\right];\label{su}\\
|g,\sigma\rangle&\equiv&|^1{\rm S}_0,1/2,\sigma\rangle;\label{sg}\\
|c,\xi\rangle&\equiv&|^3{\rm P}_2,3/2,\xi\rangle,\label{sc}\\
&&(\sigma=\uparrow,\downarrow,\xi=\pm 3/2).\nonumber
\end{eqnarray}
where the states $|g,\sigma\rangle$ and $|c,\xi\rangle$ ($\sigma=\uparrow,\downarrow,\xi=\pm 3/2$) have the same definitions as in the above two sections. In addition, the eigen energies of $H_{\rm 1b}$ corresponding to the states in Eqs.~(\ref{sd},\ref{su},\ref{sg},\ref{sc}) can be denoted as ${\cal E}_{d}$, ${\cal E}_{u}$, ${\cal E}_{g}$ and ${\cal E}_{c}$, respectively, and are all independent  of the values of $\sigma$ or $\xi$. Namely, these eigen states are two-fold degenerate. Moreover, the energy gaps between the states $|d,\sigma\rangle$, $|u,\sigma\rangle$, and $|c,\xi\rangle$ $(\sigma=\uparrow,\downarrow,\xi=\pm 3/2)$ are (Appendix~\ref{mm2}, Fig.~\ref{scheme}(b)):
\begin{eqnarray}
{\cal E}_{u}-{\cal E}_{d}&=&\Omega_{\rm eff};\label{eeu}\\
{\cal E}_{c}-{\cal E}_{d}&=&E^{\rm (AC)}_{-3/2}-E^{\rm (AC)}_{\uparrow}+\Omega_{\rm eff}/2\equiv \delta_{cd},\label{1beg}
\end{eqnarray}
where $\Omega_{\rm eff}>0$ is the Rabi frequency of the Raman coupling and $E^{\rm (AC)}_{\zeta}$ ($\zeta=\downarrow,3/2$) is the AC-Stark shift of the state $|^3{\rm P}_2,3/2,\zeta\rangle$, which can be controlled by the intensities of the Raman  beams. Our current scheme works in the region with $\delta_{cd}>0$.

\begin{figure}
\includegraphics[width=0.38\textwidth]{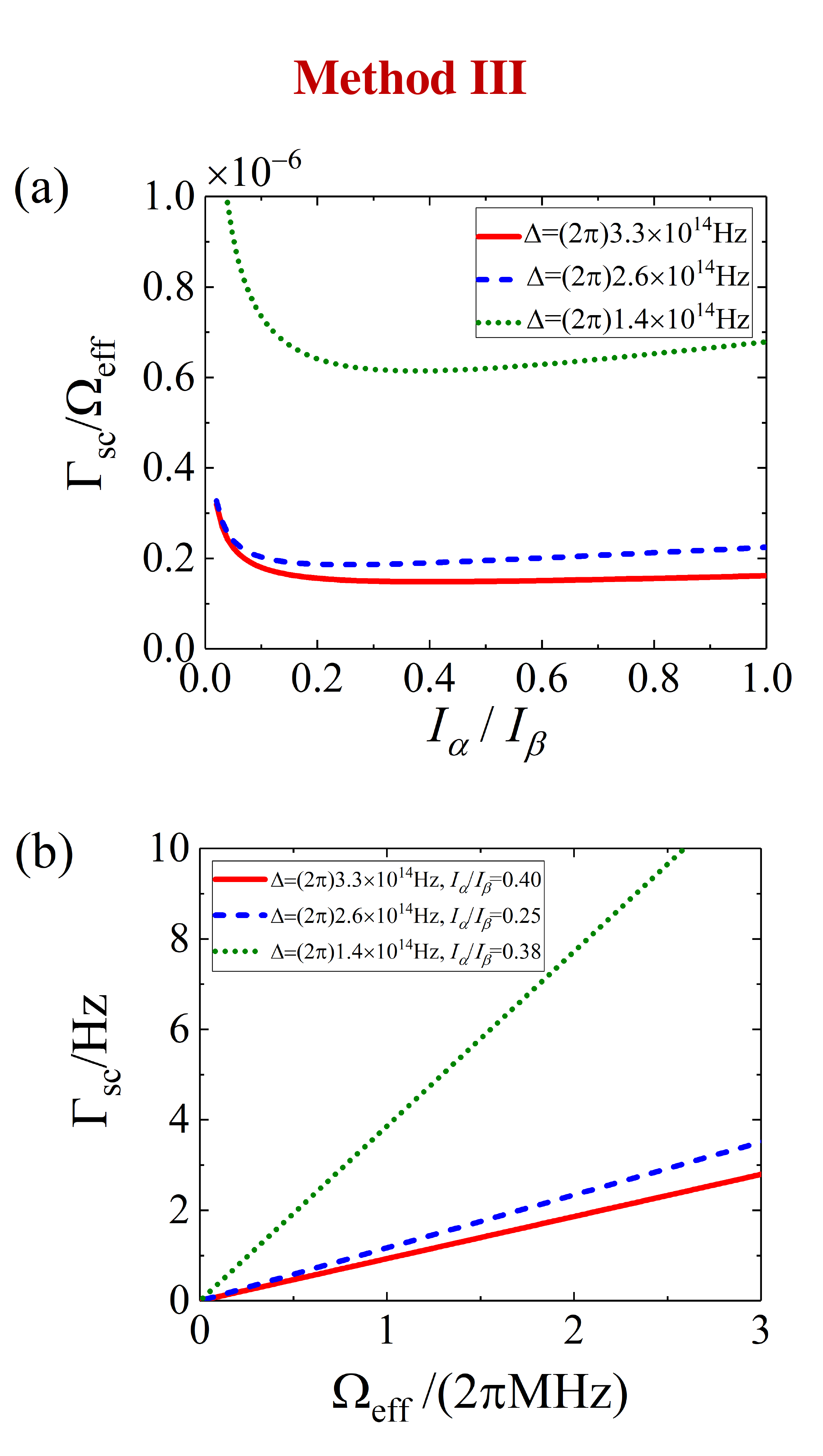}
\caption{(color  online) {\bf (a)} The ratio  between  the photon scattering rate $\Gamma_{\rm sc}$ and the effective Rabi frequency $\Omega_{\rm eff}$ of the system of method III, as a function of the laser intensity ratio $I_\alpha/I_\beta$ for the cases
where the one-photon detuning $\Delta$ the Raman beams with respect to the $^3$S$_1$ state (Fig.~\ref{scheme}(a)) takes the values
 $\Delta=(2\pi)3.3\times 10^{14}$Hz (red solid line), $\Delta=(2\pi)2.6\times 10^{14}$Hz (blue dashed line), and $\Delta=(2\pi)1.4\times 10^{14}$Hz (green dotted line).  {\bf (b)} $\Gamma_{\rm sc}$ as a function of  $\Omega_{\rm eff}$ for  the cases of (a). For each $\Delta$ the laser intensity ratio $I_\alpha/I_\beta$ takes the value to minimize $\Gamma_{\rm sc}/\Omega_{\rm eff}$.}
\label{heating}
\end{figure}

Furthermore, this $^3{\rm P}_2$-$^3{\rm P}_0$ Raman coupling is much stronger than the Raman coupling between different hyperfine states  of an ultracold alkali atom, or equivalently, the heating effect for our system  is much lower than the one for the alkali atoms.
This can be understood with the following analysis.
As  in Sec.~\ref{m1a} and Sec.~\ref{nm2}, our current method works in the large-detuning cases where the detuning of the one-photon transition induced by each Raman laser is much larger than the fine splitting of the corresponding excited state.
For an  alkali atom in the electronic ground state,  the two hyperfine levels coupled by the Raman beams are in orthogonal atomic spin states. As a result, this Raman coupling cannot be created only by the laser-induced electric dipole transition (EDT). It is essentially led by both the EDT and the spin-orbit coupling of the atomic excited states.
Thus, the Raman coupling is very weak for the large-detuning cases. However, for our system, the $^3{\rm P}_0$ and $^3{\rm P}_2$ states coupled by the Raman beams have a non-zero probability to be in the same state of electronic and nuclear spin. Therefore, the Raman effect can be induced only by EDT. Thus,
one can obtain  strong Raman coupling   in the large-detuning cases.

Explicitly,  for our system the ratio $\Gamma_{\rm sc}/\Omega_{\rm eff}$ between  the photon scattering rate $\Gamma_{\rm sc}$ and the effective Rabi frequency $\Omega_{\rm eff}$ of the Raman coupling is a function of
$\Delta$ and $I_\alpha/I_\beta$, where
$\Delta$  is the one-photon detuning the Raman beams with respect to the $^3$S$_1$ state (Fig.~\ref{scheme}(a)), and $I_{\alpha,\beta}$ are the intensities of the two Raman beams $\alpha$ and $\beta$ (we denote $\alpha$ as the beam with higher frequency, as shown in Fig.~\ref{scheme}(a)). In Fig.~\ref{heating}(a) we illustrate the variation of $\Gamma_{\rm sc}/\Omega_{\rm eff}$
with $I_\alpha/I_\beta$,
for various $\Delta$, which are calculated in Appendix~\ref{mm2}. It is shown that for a fixed $\Delta$ the ratio $\Gamma_{\rm sc}/\Omega_{\rm eff}$  can be minimized when $I_\alpha/I_\beta$ is tuned to a particular value.
In Fig.~\ref{heating}(b) we further show $\Gamma_{\rm sc}$ as a function of $\Omega_{\rm eff}$, with $I_\alpha/I_\beta$ taking the value to minimize $\Gamma_{\rm sc}/\Omega_{\rm eff}$. It is shown that
when $\Delta$ is as large as $(2\pi)3.3\times 10^{14}$Hz, one can realize a Raman coupling with  $\Omega_{\rm eff}$ being  several $(2\pi)$MHz, while  the photon scattering rate $\Gamma_{\rm sc}$ is still of the order of Hz.

\subsection {Effective Inter-Atomic  Interaction}
\label{ss}

Our current scheme is to control the SEI of two atoms being in the $^1$S$_0$ state and  the lower $^3$P dressed state ($d$-state), respectively, as shown in Fig.~\ref{scheme}(c). Now we derive the effective interaction between these two atoms. Since the analysis is very similar to the one of Sec.~\ref{ss1}, here we only show the main results.

We consider the $s$-wave  scattering of these two atoms in the zero-energy limit, which is at a zero magnetic field. For this scattering process, there are the following  open channels:
\begin{eqnarray}
|g,\sigma;d,\sigma^\prime\rangle&\equiv&\frac{1}{\sqrt{2}}\left[
|g,\sigma\rangle_1|d,\sigma^\prime\rangle_2-|d,\sigma^\prime\rangle_1|g,\sigma\rangle_2
\right],\nonumber\\
&&\hspace{3cm}(\sigma,\sigma^\prime=\uparrow,\downarrow),\label{gj}
\end{eqnarray}
as well as the closed channels:
\begin{eqnarray}
|g,\sigma;u,\sigma^\prime\rangle&\equiv&\frac{1}{\sqrt{2}}\left[
|g,\sigma\rangle_1|u,\sigma^\prime\rangle_2-|u,\sigma^\prime\rangle_1|g,\sigma\rangle_2
\right];\label{gu}\\
|g,\sigma;c,\xi\rangle&\equiv&\frac{1}{\sqrt{2}}\left[
|g,\sigma\rangle_1|c,\xi\rangle_2-|c,\xi\rangle_1|g,\sigma\rangle_2
\right],\nonumber\\
&&(\sigma,\sigma^\prime=\uparrow,\downarrow;\ \ \ \xi=\pm 3/2).\label{gc}
\end{eqnarray}
The energy gap between the open channel  and the closed channels in Eq.~(\ref{gu}) and Eq.~(\ref{gc}) are $\Omega_{\rm eff}$ and $\delta_{cd}$, respectively, as shown in
Fig.~\ref{scheme}(d).
In addition,
for our system, the inter-atomic interaction ${\hat V}({\bf r})$
can be  expressed as
\begin{eqnarray}
{\hat V}({\bf r})={\hat V}^{(0)}(r)+{\hat V}^{(2)}({\bf r}),
\end{eqnarray}
where ${\hat V}^{(0(2))}(r)$  is the interaction between two atoms in the $^1$S$_0$ and $^3$P$_{0}$ ($^3$P$_{2}(F=3/2)$) states, respectively, with the explicit forms being given  in Appendix~\ref{int}.

\begin{figure*}
\includegraphics[width=0.7\textwidth]{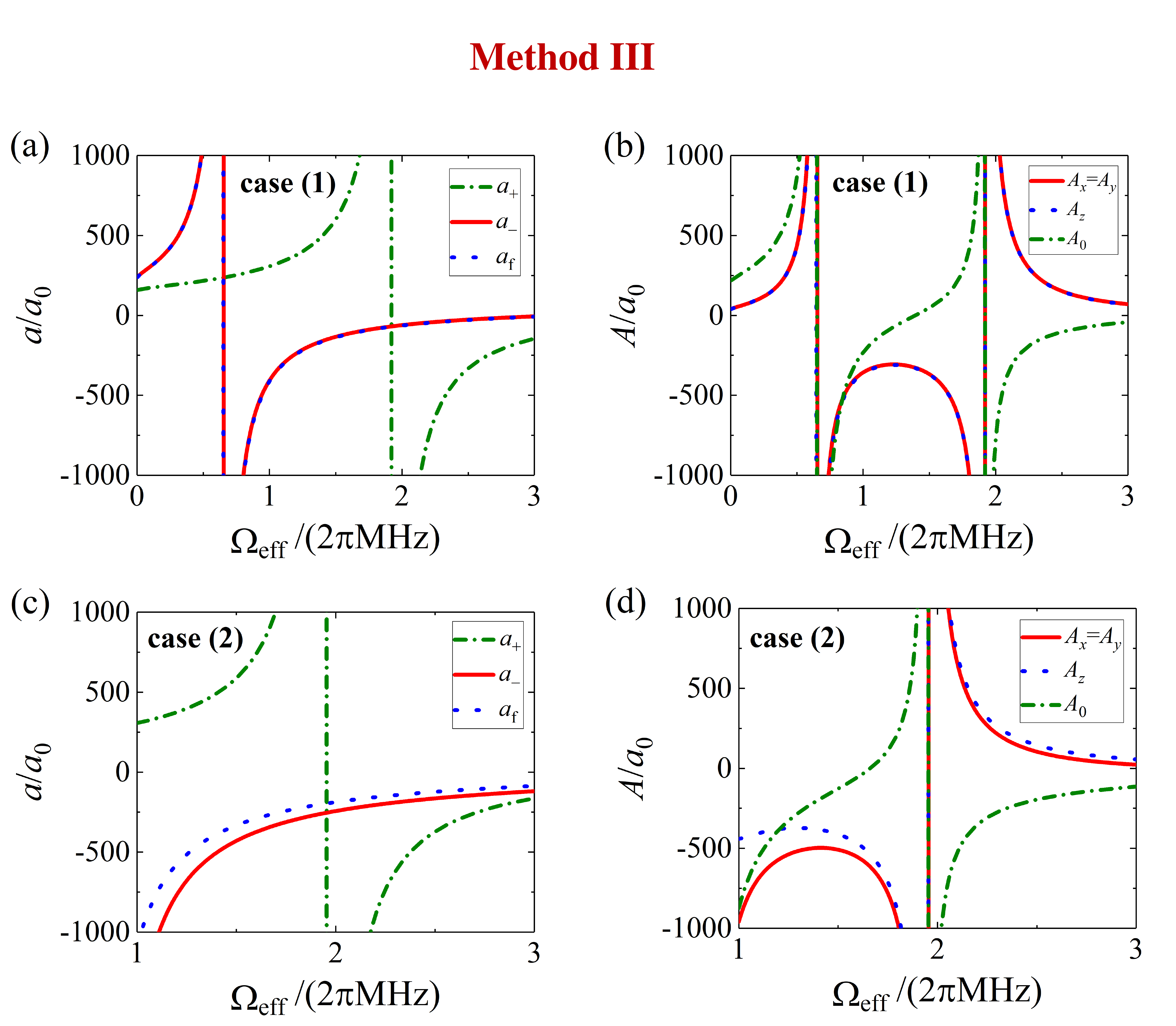}
\caption{(color  online) The scattering lengths $a_{\pm,{\rm f}}$ ({(a)}, {(c)}) and interaction parameters $A_{x,y,z,0}$ ({(b)}, {(d)}) of the system of method III. Here we show the results for two cases, which are given by the multi-channel square-well model  in Sec.~\ref{model1},  with width $b=81a_0$ and other parameters being given in Table III of Appendix~\ref{apppara}. The values of $\Omega_{\rm eff}$ under which we have  $A_0=0$, as well as the corresponding values of  $A_{x,y,z}$ and $a_{\pm, f}$, are shown in Appendix~\ref{appd}.}
\label{sw2}
\end{figure*}

As in Sec.~\ref{ss1}, a straightforward analysis based on the form of ${\hat V}({\bf r})$ (Appendix~\ref{int}) shows that   in the zero-energy limit if the two atoms were incident from one of the following four states, i.e., the polarized states $|g,\uparrow;d,\uparrow\rangle$ and $|g,\downarrow;d,\downarrow\rangle$ as well as the states $|\pm \rangle$ defined by
\begin{eqnarray}
|\pm\rangle\equiv\frac{1}{\sqrt{2}}\big[|g,\uparrow;d,\downarrow\rangle\mp |g,\downarrow;d,\uparrow\rangle\big],\label{pm}
\end{eqnarray}
then the output state of the scattering process would be exactly the same as the incident state, i.e.,
 there is only elastic scattering. We denote the scattering lengths corresponding to the incident states $|\pm \rangle$ as $a_{\pm}$. In addition, due to the reflection symmetry with respect to the $x-y$ plane, the scattering lengths corresponding to the incident states $|g,\uparrow;d,\uparrow\rangle$ and $|g,\downarrow;d,\downarrow\rangle$ are the same, and can be denoted as $a_{\rm f}$. Thus, the amplitude for the spin-exchange process
$
|g,\uparrow;d,\downarrow\rangle\Leftrightarrow |g,\downarrow;d,\uparrow\rangle\label{ie}
$
is just $(a_--a_+)/2$.

Therefore, the low-energy interaction between these two atoms can be described
by the pseudo potential
\begin{eqnarray}
&&{\hat V}_{\rm eff}=\frac{2\pi}{\mu}\Big[a_{\rm +}|+\rangle\langle+|+a_{-}|-\rangle\langle-|+a_{\rm f}{\hat P}_{\rm f}\Big]\delta({\bf r})\frac{\partial}{\partial r}(r\cdot),\nonumber\\
\label{veff}
\end{eqnarray}
where
\begin{eqnarray}
{\hat P}_{\rm f}=\sum_{\sigma=\uparrow,\downarrow}|g,\sigma;d,\sigma\rangle\langle g,\sigma;d,\sigma|
\end{eqnarray}
is the projection operator of the polarized states.
As in the above sections, we can
treat the atoms in $|g,\uparrow\!(\downarrow)\rangle$ and $|d,\uparrow(\downarrow)\rangle$ as  two distinguished particles with pseudo-spin $1/2$, and express the effective two-atom Hamiltonian as
\begin{eqnarray}
H_{\rm 2body}^{\rm (eff)}=\frac{{\bm p}_S^2}{2m}+\frac{{\bm p}_P^2}{2m}+{\hat V}_{\rm eff}, \nonumber
\end{eqnarray}
with
${\bm p}_{S(P)}$ being the momentum operator  of the $g$- ($d$-) atom and
${{\bm p}_S^2}/(2m)+{{\bm p}_P^2}/{(2m)}$ being the {\it pseudo-spin independent free Hamiltonian}, and
\begin{eqnarray}
&&{\hat V}_{\rm eff}=\frac{2\pi}{\mu}\times\nonumber\\
&&\left[\!\frac{A_{x}}2\hat{\sigma}_x^{(S)}\hat{\sigma}_x^{(P)}\!
+\!\frac{A_{y}}2\hat{\sigma}_y^{(S)}\hat{\sigma}_y^{(P)}\!
+\!\frac{A_{z}}2\hat{\sigma}_z^{(S)}\hat{\sigma}_z^{(P)}\!+\!A_0
\right]\!
\delta({\bf r})\frac{\partial}{\partial r}(r\cdot)\nonumber\\
\label{aeff2}
\end{eqnarray}
is the effective inter-atomic interaction, which is equivalent to the one of Eq.~(\ref{veff}).
For the current system, ${\hat{\sigma}}_{\pm}^{(j)}=({\hat{\sigma}}_{x}^{(j)}\pm i{\hat{\sigma}}_{y}^{(j)})/2$
 ($j=S,P$), and ${\hat{\sigma}}_{x,y,z}^{(S(P))}$ are the Pauli operators of the pseudo spin of the $g$-atom ($d$-atom). In addition, the relation between coefficients $A_{x,y,z,0}$ and the scattering lengths $a_{\pm,{\rm f}}$ are same as Eqs.~(\ref{aex2}-\ref{a02}), i.e.,
\begin{eqnarray}
A_{x}&=&A_y\nonumber\\
&=&\frac{a_--a_{+}}{2};\label{aex}\\
A_z&=&\frac{2a_{\rm f}-(a_-+a_+)}2;\\
A_0&=&\frac{2a_{\rm f}+(a_-+a_+)}4.\label{a0}
\end{eqnarray}

\subsection{Resonant Control of ${\hat V}_{\rm eff}$}

Due to the conservation of the angular momentum component $M$ defined in Eq.~(\ref{bigM}), the  $s$-wave states of open channels of our system are coupled to both the $s$-wave  and the  $d$-wave states of the closed channels
%states of the closed channels corresponding to $|g,\uparrow;u,\downarrow\rangle$ and $|g,\downarrow;u,\uparrow\rangle$, as well as the $d$-wave states of the closed channels corresponding to $|g,\uparrow;u,\uparrow\rangle$, $|g,\downarrow;u,\downarrow\rangle$, and $|g,\sigma;c,\xi\rangle$ ($\sigma=\uparrow,\downarrow, \xi=\pm1/2$),
by the interaction ${\hat V}({\bf r})$. Furthermore,
the energy gap $\delta_{cd}$ ($\Omega_{\rm eff}$)
between the closed channels $|g,\sigma,c,\xi\rangle$ ($|g,\sigma,u,\sigma^\prime\rangle$) and the open channels $|g,\sigma,d,\sigma^\prime\rangle$
($\sigma,\sigma^\prime=\uparrow,\downarrow,\xi=\pm 3/2$) (Fig.~\ref{scheme}(d))
can be of the same order of magnitude with the der Waals energy scale $E_{\rm vdW}$ of Yb atom ($(2\pi)$MHz) with a low heating rate.
Therefore,
as discussed in Sec.~\ref{introduction}, it is very possible that by tuning $\Omega_{\rm eff}$ and $\delta_{cd}$, one can
make the threshold of the open channels to be near resonant to a closed-channel bound state,
 and thus realize a
Feshbach resonance, while keeping the heating rate to be low-enough.
The interaction parameters $a_{\pm,{\rm f}}$ or $A_{x,y,z,0}$ can be efficiently manipulated via these resonances.
%even if the intra-channel potential of the closed channels are not at resonance.

%and the closed channels of our current system are much more than the ones of  methods I and II, according to

In addition, as mentioned before, so far we assume the frequency difference of these two Raman beams takes a certain value to compensate for the AC-Stark shift difference of the $^3$P$_0$ and $^3$P$_2$ states.
It is clear that  this frequency difference can be tuned to other values. Thus, if it is required, one can use both laser intensity and this frequency difference as control parameters for inter-atomic interaction.

%is {\it spin-permitted}, which occurs without the help of the fine splitting  of the excited state. That is very different from the Raman coupling between different hyperfine states in the electronic ground manifold of an alkaline atom, which cannot occur without the electronic SOC,

%and is more like the AC Stark shift or the optical confinement potential of a single internal state of an ultracold atom. Thus, the heating effect of our Raman coupling  is much weaker than the former one and similar to the latter one. Accordingly, the effective Rabi frequency $\Omega_{\rm eff}$ of our Raman coupling can be much stronger than the one of the Raman beams for ultracold alkaline atoms, and can be at most as large as the depth of the optical trapping potential.

\subsection{Illustration with multi-channel square-well model}
\label{model2}

As in Sec.~\ref{model1},
we illustrate our current scheme with the multi-channel square-well model. Explicitly,
the model for the potential ${\hat V}^{(2)}({\bf r})$ is the same as in Sec.~\ref{model1}. In addition, as shown in Appendix~\ref{int},
 the $^1$S$_0$-$^3$P$_0$ interaction potential ${\hat V}^{(0)}(r)$ can be formally expressed as ${\hat V}^{(0)}(r)=\sum_{j=+,-}V^{(0)}_{j}(r){\hat P}^{(0)}_j$, where ${\hat P}^{(0)}_\pm$ is the projection operator for the  electronic  state $|\psi_{\pm}\rangle$ defined in Appendix~\ref{int}, and $V^{(0)}_{\pm}(r)$ are the corresponding interaction potentials. In our calculation, we model $V^{(0)}_{\pm}(r)$ also as square-wells, i.e.,
\begin{eqnarray}
V^{(0)}_{\pm}(r)=-U^{(0)}_{\pm}\theta(b-r);\ \ (r\geq 0),
\end{eqnarray}
where  $\theta(x)$ is the step function as before. In our calculation, we choose the range $b$  of all the square-well potentials $V^{(0)}_{\pm}(r)$ and $V^{(2)}_{1,...,6}(r)$  as $81a_0$.
We display the results for two typical cases in Fig.~\ref{sw2}. The parameters for these cases are given in Table III of Appendix~\ref{apppara}.
Fig.~\ref{sw2} shows that the scattering lengths $a_{\pm}$ and $a_{\rm f}$ or the interaction parameters $A_{x}=A_y$ and
 $A_{z,0}$ can be efficiently manipulated via the resonances induced by the variation of $\Omega_{\rm eff}$.
%It is shown that, similar as in Sec.~\ref{model1}, resonances occurs when then  effective Rabi frequency $\Omega_{\rm eff}$ is of the order of $(2\pi)$MHz {\color{red}, even if the background scattering lengths are....}. Using these resonances one can efficiently manipulate the scattering lengths $a_{\pm,{\rm f}}$ or the parameters $A_{{\rm ex},z,0}$ of the effective interaction ${\hat V}_{\rm eff}$ and tune the system to various parameter regions, {\it e.g.}, the ``anti-ferromagnetic-like" or  ``ferromagnetic-like" regions defined in Sec.~\ref{model1}.

\section{Discussions}
\label{summary}

In this work, we propose three methods for the laser manipulation of the SEI between two ultracold fermionic alkaline-earth (like) atoms in electronic  $^1$S$_0$ state  and $^3$P$_2$ state (or  $^3$P$_2$-$^3$P$_0$ dressed state), respectively. Our methods are based on the spin-dependent AC-Stark effect of the $^3$P$_2$ states, or the $^3$P$_2$-$^3$P$_0$ Raman coupling. We show that  the laser-induced heating corresponding to both of these two effects is very weak. By tuning the AC-Stark shift difference $\Delta_{\rm AC}$ or the effective Rabi frequency $\Omega_{\rm eff}$ one can induce Feshbach resonances with which the SEI can be efficiently controlled.
In particular, for the systems of methods II and III the appearance of the ``low-heating" Feshbach resonances is quite possible for realistic systems, and does not require the scattering lengths of the bare inter-atomic interaction potentials ${\hat V}^{(0)}(r)$ and ${\hat V}^{(2)}({\bf r})$ to be very large. For instance,
as shown in Tables II and III of Appendix~\ref{apppara},
in our calculations with multi-channel square-well models for these two methods for $^{171}$Yb atoms, we set the scattering lengths $a_{\pm}^{(0)}$ for the potential ${\hat V}^{(0)}(r)$ to be $232a_0$ and $372a_0$, which are reported by the references \cite{Oscar2003,Ono2019,Abeln2021,DaWu2022}, and set all the other scattering lengths in our interaction model to be less than $200a_0$. As illustrated in Fig.~\ref{rsw2} and Fig.~\ref{sw2}, Feshbach resonances with low heating rates can appear for these cases.

In the above sections, we take ultracold $^{171}$Yb atoms as an example. Our methods are also possible to be applicable for other types of fermionic alkaline-earth (like) atoms, {\it e.g.}, $^{173}$Yb \cite{PhysRevA.89.031601} or $^{87}$Sr atoms \cite{PhysRevA.99.052503}.

At the end of this paper, we give the following  comments for these methods:

(I): Our above analysis as well as the illustrations with the multi-channel square-well models just show it is quite possible to realize Feshbach resonances with the laser-induced heating rate being of the order of Hz or even lower.
Nevertheless, for realistic systems there does exist the possibility that the resonances only occur in the regions with the heating rate being larger than $10$Hz, since sometimes the binding energy of the shallowest bound state of a van der Waals interaction potential can be larger than  $E_{\rm vdW}$ by one order of magnitude.
As mentioned above,
unfortunately, we cannot make quantitative predictions for realistic systems with specific atoms,
due to the  lack of detailed parameters of the  interaction potential. The positions and widths of the Feshbach resonances for specific atoms, as well as the significance of the laser-induced heating in the resonance region, should be examined via experiments or multi-channel numerical calculations with accurate inter-atomic interaction potentials.
(Notice that  our system is complicated because there are many degenerate closed channels that are coupled with each other. Thus, for a specific type of atom one cannot predict if  resonances can appear for a certain parameter region ({\it e.g.}, the regions with  $\Delta_{\rm AC}<(2\pi)10$MHz   or  $\Omega_{\rm eff}<(2\pi)10$MHz) even with the analysis based on the theory of single-channel van der Waals potential.)

%In particular, for the square-well models we do find some cases in which there is no Feshbach resonance when $\Delta_{\rm AC}<(2\pi)10$MHz (method I) or $\Omega_{\rm eff}<(2\pi)10$MHz (method II). These results show that we cannot guarantee the appearance of the resonances when $\Delta_{\rm AC}$ or $\Omega_{\rm eff}$ takes such low values.

(II): Here we can make a brief comparison for these three methods. Method I is the most simple one because only one laser beam is required. Method II is a little bit complicated because it requires two laser beams. Nevertheless, as shown in Sec.~\ref{nm2}, the frequency difference of these two beams is not required to be locked to a certain value. Thus, this method is still easier to be realized than the usual Raman schemes. Method III is the most complicated one because the atoms are in a superposition state of $^3$P$_2$ and $^3$P$_0$ levels, and the frequency difference of the two Raman beams should be fixed. Fortunately, since the effective Rabi frequency $\Omega_{\rm eff}$ of our system is as large as $(2\pi)$MHz, the fluctuation of this frequency difference is required only to be much less than this order. This requirement can be realized in current experiments.

Moreover, in methods II and III the $^3$P atoms should be prepared in the lower dressed states $|q,\sigma\rangle$ or $|d,\sigma\rangle$ ($\sigma=\uparrow, \downarrow$).
The  preparation and detection of the dressed states may induce
imperfections and complications for the experiments. More discussions for the detection of the dressed states are given in the following point (III).

On the other hand, by comparing Fig.~\ref{heating1beam}, Fig.~\ref{hh2}, and  Fig.~\ref{heating} we find that, for the realization of a fixed energy gap between the open and closed channels via lasers with given one-photon detuning, the photon scattering rate $\Gamma_{\rm sc}$ of the laser beams of method III is usually lower than the ones of methods I and II. This result implies that  method III has the weakest heating effect. This is an important advantage of this method.

Finally, as  discussed before, the possibility for the appearance of Feshbach resonances  with a low-enough heating rate
is higher for the systems with method II and III, due to the coupling between $s$-wave states of the open and closed channels.
In addition, since the potential curves involved in the closed channels of these three methods are quite different, for a specific system it is possible that   this kind of resonances cannot be realized via one method, but can be realized via another method.

(III) Here we discuss the measurement of the dressed states of methods II and III.
Due to the energy conservation, after the scattering processes the atoms would return to the open channels with
 lower dressed states $|q,\uparrow\!(\downarrow)\rangle$ (method II) or $|d,\uparrow\!(\downarrow)\rangle$ (method III).
 Therefore, in most cases, only these lower dressed states should be detected.
According to Eqs. (\ref{hq1}, \ref{hq2}), $|q,\uparrow\rangle$ is the superposition of the $^3$P$_2(F=3/2)$ states with $m_F=-1/2$ and $+3/2$, while $|q,\downarrow\rangle$ is the superposition of states with $m_F=1/2$ and $-3/2$. Therefore, one can use a Stern-Gerlach experiment to detect the number of atoms for each $m_F$, and then derive the populations of $|q,\uparrow\rangle$ and  $|q,\downarrow\rangle$ from these atom numbers. Similarly,  since the states $|d,\uparrow\!(\downarrow)\rangle$ are the $^3$P$_2$-$^3$P$_0$ dressed states with $m_F=-1/2(+1/2)$, one can use a Stern-Gerlach experiment to detect the atom numbers for the
$^3$P$_2(m_F=-1/2)$ and  $^3$P$_2(m_F=+1/2)$ states, and then derive the populations of $|d,\uparrow\rangle$ and $|d,\downarrow\rangle$ from these atom numbers.

(IV): In this work we perform the calculations in the zero-energy limit. For realistic systems,  one may require to consider the effect induced by the finite incident momentum $k$. When $k$ is finite there may be new spin-change scattering processes, {\it e.g.}, the process $|g,\uparrow;\xi,\uparrow\rangle\Leftrightarrow(|g,\uparrow;\xi,\downarrow\rangle+|g,\downarrow;\xi,\uparrow\rangle)/\sqrt{2}$, with $\xi=c$ for method I and $\xi=d$ for method III. These processes are induced by the fact that the interaction potential $V^{(2)}({\bf r})$ is anisotropic. However,  the scattering amplitudes of these  processes are proportional to $k^2$ and thus can be ignored when $k$ is low enough, because the $d$-wave motional states are involved. More importantly, direct analysis based on the symmetry of the interaction potential shows that the interaction potential can never couple the singlet state to the triplet states of the two-atom pseudo-spin we defined before,
even in the finite-$k$ cases. Therefore total pseudo-spin of the two atoms is always conserved, and the singlet state is always a non-degenerate eigen state of the effective inter-atomic interaction.

\begin{acknowledgments}
We thank Prof. Y. Takahashi, Prof. Ran Qi, Prof. Meng Khoon Tey, Prof. Jinyi Zhang, Prof. Xibo Zhang,  Prof. Ren Zhang, and Dr. Da-Wu Xiao for very helpful discussions.
This work is supported in part by the National Key Research and
Development Program of China Grant No. 2018YFA0306502, NSAF (Grant No.
U1930201), as well as the Research Funds of Renmin University of China
under Grant No. 16XNLQ03.
\end{acknowledgments}

\appendix

\begin{widetext}

\section{Single-Atom Effective Hamiltonian and Heating Effects}

In this appendix, we derive the single-atom effective Hamiltonian and its eigen energies and eigen states, as well as the laser-induced heating rates or the photon scattering rates for the systems of methods I, II, and III.

\subsection{The System of Method I}
\label{a1}

In this section, we derive the
AC-Stark shifts $E_{\eta}^{\rm (AC)}$ ($\eta=\uparrow,\downarrow,\pm 3/2$), the AC-Stark shift difference $\Delta_{\rm AC}$, and the photon scattering rate $\Gamma_{\rm sc}$ for the system of Sec.~\ref{m1}.
Here we take the $^3$P$_2(F=3/2)$ states with magnetic quantum numbers $-1/2$ ($\uparrow$)
and $-3/2$
as an example.   The states with magnetic quantum numbers $1/2$ ($\downarrow$)
and $+3/2$
have the same effective Hamiltonian and heating rate.
Accordingly,  in the following calculation, we  consider the  $^3$P$_2$ states
\begin{eqnarray}
 |c,\uparrow\rangle&\equiv&|^3{\rm P}_2,3/2,-1/2\rangle=\sqrt{\frac 25}|^3{\rm P}_2,m_J=0\rangle_{ele}|-1/2\rangle_{nuc}-\sqrt{\frac 35}|^3{\rm P}_2,m_J=-1\rangle_{ele}|+
 1/2\rangle_{nuc};\\
 \nonumber\\
 |c,-3/2\rangle&\equiv&|^3{\rm P}_2,3/2,-3/2\rangle=|^3{\rm P}_2,m_J=-1\rangle_{ele}|-1/2\rangle_{nuc},
\end{eqnarray}
as well as the $^3{\rm S}_1$ and $^3{\rm D}_{1,2,3}$ states which can be coupled to $|c,\uparrow\rangle$ and $|c,-3/2\rangle$  via the $\pi$-polarized beam, i.e.,
\begin{eqnarray}
 |f_1\rangle&\equiv&|^3{\rm S}_1,m_J=0\rangle_{ele}|-1/2\rangle_{nuc};\hspace{1.33cm}
 |f_2\rangle\equiv|^3{\rm D}_1,m_J=0\rangle_{ele}|-1/2\rangle_{nuc}; \label{f1}\\
 |f_3\rangle&\equiv&|^3{\rm D}_2,m_J=0\rangle_{ele}|-1/2\rangle_{nuc};\hspace{1.26cm}
 |f_4\rangle\equiv|^3{\rm D}_3,m_J=0\rangle_{ele}|-1/2\rangle_{nuc};\\
  |f_5\rangle&\equiv&|^3{\rm S}_1,m_J=-1\rangle_{ele}|+1/2\rangle_{nuc};\hspace{1.07cm}
|f_6\rangle\equiv|^3{\rm D}_1,m_J=-1\rangle_{ele}|+1/2\rangle_{nuc};\\
  |f_7\rangle&\equiv&|^3{\rm D}_2,m_J=-1\rangle_{ele}|+1/2\rangle_{nuc};\hspace{1cm}
 |f_8\rangle\equiv|^3{\rm D}_3,m_J=-1\rangle_{ele}|+1/2\rangle_{nuc};\\
 |f_9\rangle&\equiv&|^3{\rm S}_3,m_J=-1\rangle_{ele}|-1/2\rangle_{nuc};\hspace{0.9cm}
 |f_{10}\rangle\equiv|^3{\rm D}_1,m_J=-1\rangle_{ele}|-1/2\rangle_{nuc};\\
  |f_{11}\rangle&\equiv&|^3{\rm D}_2,m_J=-1\rangle_{ele}|-1/2\rangle_{nuc};\hspace{0.85cm}
 |f_{12}\rangle\equiv|^3{\rm D}_3,m_J=-1\rangle_{ele}|-1/2\rangle_{nuc}.\label{f12}
\end{eqnarray}
Here $|\rangle_{ele}$ is the electronic state, with $m_J$ being the magnetic quantum number of  electronic total angular momentum ({\it e.g.,} $|^3{\rm S}_1,m_J=0\rangle_{ele}$ is the $^3{\rm S}_1$ state with $m_J=0$), and $|\sigma\rangle_{nuc}$ ($\sigma=\pm 1/2$) is the nuclear spin state with magnetic quantum number $\sigma$.
Here the excited states $|f_j\rangle$  ($j=1,...,12$) are expressed as direct products of electronic and nuclear spin states. That is  because the hyperfine splitting of $^3{\rm S}_1$ and $^3{\rm D}_{1,2,3}$ levels can be ignored for our system, as mentioned in Sec.~\ref{m1a}. In addition, according to the selection rule, the states $|f_{1,...,8}\rangle$ are coupled only to
 $|c,\uparrow\rangle$ by the laser beam, while $|f_{9,...,12}\rangle$ are coupled only to  $|c,-3/2\rangle$.

The single-atom Hamiltonian for our system can be expressed as ($\hbar=1$):
\begin{eqnarray}
H_S&=&\sum_{\eta=\uparrow,-3/2}E_\eta|c,\eta\rangle\langle c,\eta|
+\sum_{j=1}^{12} \left(E_j-i\gamma_j/2\right)|f_j\rangle\langle f_j|
+\sum_{\eta=\uparrow,-3/2}\sum_{j=1}^{12}
\Big[\Omega_{j\eta}\cos(\omega t) |c_j\rangle\langle c, q|+h.c.\Big],
%\nonumber\\
%&&+\sum_{j=1}^8
%\left\{\left[\Omega_{jb}^{(\alpha)}\cos\omega_\alpha t+\Omega_{jb}^{(\beta)}\cos\omega_\beta t\right]|c_j\rangle{\color{red} \langle c,\uparrow|}+h.c.\right\},
\label{hs1}
\end{eqnarray}
where $\omega$ is the angular frequency of the laser beam, and we do not make the rotating-wave approximation.
Here $E_{\eta}$ ($\eta=\uparrow,-3/2$) is the energy of state $|c,\eta\rangle$, while $E_{j}$  and $\gamma_j$
($j=1,...,12$)
are the energy and spontaneous emission rate
of excited state $|f_j\rangle$. According to the above definitions, we have
$E_j=E_{j+4}$ and $\gamma_{j}=\gamma_{j+4}$ ($j\leq 8$). In Eq.~(\ref{hs1}), $\Omega_{j\eta}$ ($\eta=\uparrow,-3/2;\ j=1,...,12$) is the Rabi frequency of the laser-induced coupling between states $|c,\eta\rangle$ and $|f_j\rangle$, which  can be further expressed as
 \begin{eqnarray}
\Omega_{j\eta}=\sqrt{\frac{2}{\epsilon_0c}}
\langle f_j|D_z|c,\eta\rangle\sqrt{I};\ \ \ (\eta=\uparrow,-3/2;\ \  j=1,...,12),
 \end{eqnarray}
 where $\epsilon_0$ and $c$ are the vacuum dielectric constant and speed of light, respectively, $I$ is the intensity of the laser beam, and  $\langle f_j|D_z|c,\eta\rangle$ is the matrix element of the atomic electric-dipole along the $z$-direction, with respect to the states $|f_j\rangle$ and $|c,\eta\rangle$. Without loss of generality, we assume $\langle f_j|D_z|c,\eta\rangle$ is real. As mentioned above,  the selection rules yield
$
\langle f_{j}|D_z|c,\uparrow\rangle=0$;
for $j=9,...,12$
and
$
\langle f_{j}|D_z|c,-3/2\rangle=0
$
for $j=1,...,8$.
 In this work, we derive the values of $\langle f_{j}|D_z|c,\eta\rangle$ ($\eta=\uparrow,-3/2;\ \  j=1,...,12$) via the Wigner–Eckart theorem, with the corresponding reduced matrix element given by Ref~\cite{Porsev1999} for $^{171}$Yb. In addition,  in Eq.~(\ref{hs}) we phenomenologically describe the spontaneous emission of the excited states via the non-Hermitian  term proportional to $i\gamma_j$ ($j=1,...,12$).

 Since the laser beam is far-off resonant for the direct transitions from the $^3$P states to the excited states, we can adiabatically eliminate the excited states $|f_j\rangle$ ($j=1,...,12$) and derive the effective Hamiltonian for the $^3$P states $|c,\uparrow\rangle$ and $|c,-3/2\rangle$:
 \begin{eqnarray}
H_{\rm eff}=\sum_{\eta=\uparrow,-3/2}
\left[E_{\eta}^{\rm (AC)}-i\Gamma_\eta/2\right]|c,\eta\rangle \langle c,\eta|,
\label{ar}
\end{eqnarray}
where the AC-Stark shift $E_{\eta}^{\rm (AC)}$ and the photon scattering rate $\Gamma_\eta$ for the state $|c,\eta\rangle$ ($\eta=\uparrow,-3/2$) are given by:
 \begin{eqnarray}
E_{\eta}^{\rm (AC)}&=&-\frac14\sum_{j=1}^{12}
\left[\frac{\Omega_{j\eta}^{2}}{E_j-E_\eta-\omega}
+\frac{\Omega_{j\eta}^{2}}{E_j-E_\eta+\omega}
\right],\label{eac}
\end{eqnarray}
and
\begin{eqnarray}
\Gamma_{\eta}&=&\frac14\sum_{j=1}^{12}\gamma_j
\left[\frac{\Omega_{j\eta}^{2}}{(E_j-E_\eta-\omega)^2}
+\frac{\Omega_{j\eta}^{2}}{(E_j-E_\eta+\omega)^2}
\right],
\label{arr}
\end{eqnarray}
respectively.
In the derivation of the results Eqs.~(\ref{ar}-\ref{arr}) we have also used the fact $1/(g+i\xi/2)\approx 1/g-i\xi/(2g^2)$ with $g$ and $\xi$ being real numbers and $|g|\gg |\xi|$.
Thus, the difference $\Delta_{\rm AC}$ between the AC-Stark shifts of states $|c,-3/2\rangle$ and $|c,\uparrow\rangle$ can be expressed as
\begin{eqnarray}
\Delta_{\rm AC}=E_{-3/2}^{\rm (AC)}-E_{\uparrow}^{\rm (AC)},\label{rdac}
\end{eqnarray}
as shown in Eq.~(\ref{dac}). In addition, since  the $^3$P$_2$ atoms are prepared in the state $|c,\uparrow\!(\downarrow)\rangle$, the photon scattering rate $\Gamma_{\rm sc}$ of our system is just the one for these states, i.e.,
\begin{eqnarray}
\Gamma_{\rm sc}=\Gamma_{\uparrow}.
\label{gscm1}
\end{eqnarray}
The calculations for Fig.~\ref{heating1beam} and related parts in Sec.~\ref{m1a} are based on Eqs.~(\ref{rdac}, \ref{gscm1}).

\subsection{The System of Method II}
\label{nmm2}

In this section we first derive the effective single-atom Hamiltonian for the system of Sec.~\ref{nm2}, and then calculate the eigen states $|h(q),\uparrow\rangle$ and $|h(q),\downarrow\rangle$ defined in Eqs.~(\ref{hq1}-\ref{hq4}), the eigen energy difference $\Delta_{hq}$, as well as the photon scattering rate $\Gamma_{\rm sc}$ for this system.
Since this system is a direct generalization of the system of Sec.~\ref{m1} and Appendix~\ref{a1}, our calculation is based on Appendix~\ref{a1}. For convenience, we first introduce functions $\varepsilon_{j}(I,\omega)$ and  $\lambda_{j}(I,\omega)$ $(j=1,3)$ of laser frequency $\omega$ and intensity $I$ as
\begin{eqnarray}
&&\varepsilon_{1}(I,\omega)=E_{\uparrow}^{\rm (AC)};\hspace{1cm} \varepsilon_{3}(I,\omega)=E_{-3/2}^{\rm (AC)};\\
&&\lambda_{1}(I,\omega)=\Gamma_{\uparrow};\hspace{1.5cm} \lambda_{3}(I,\omega)=\Gamma_{-3/2},
\end{eqnarray}
where $E_{\eta}^{\rm (AC)}$ and  $\Gamma_\eta$ ($\eta=\uparrow,-3/2$) are
the AC-Stark shift and  photon scattering rate defined in Eqs.~(\ref{eac}), respectively, corresponding to a $\pi$-polarized laser beam with frequency $\omega$ and intensity $I$.

In the system of  Sec.~\ref{nm2}, the two laser beams are polarized along the $x$- and $z$-directions.
As shown in Sec.~\ref{m1} and Appendix~\ref{a1}, the beam polarized along the $z$-directions can induce AC-Stark shifts for  the states
$|c,\eta\rangle$ ($\eta=\uparrow,\downarrow,\pm 3/2$ defined in Sec.~\ref{m1} (Fig.~\ref{scheme1}(a))) (i.e., the states $|^3{\rm P}_2,3/2,m_F\rangle$ ($m_F=\pm 3/2, \pm 1/2$)). Similarly, the beam polarized along the $x$-directions can induce AC-Stark shifts for the eigen states of the atomic total angular momentum  along the $x$-direction, i.e., the states:
\begin{eqnarray}
|x,-3/2\rangle&\equiv&\frac{1}{2\sqrt{2}}\left[|c,-3/2\rangle-\sqrt{3}|c,\uparrow\rangle+\sqrt{3}|c,\downarrow\rangle-|c,3/2\rangle\right];\\
|x,\uparrow\rangle&\equiv&\frac{1}{2\sqrt{2}}\left[\sqrt{3}|c,-3/2\rangle-|c,\uparrow\rangle-|c,\downarrow\rangle+\sqrt{3}|c,3/2\rangle\right];\\
|x,\downarrow\rangle&\equiv&\frac{1}{2\sqrt{2}}\left[\sqrt{3}|c,-3/2\rangle+|c,\uparrow\rangle-|c,\downarrow\rangle-\sqrt{3}|c,3/2\rangle\right];\\
|x,3/2\rangle&\equiv&\frac{1}{2\sqrt{2}}\left[|c,-3/2\rangle+\sqrt{3}|c,\uparrow\rangle+\sqrt{3}|c,\downarrow\rangle+|c,3/2\rangle\right].
\end{eqnarray}
Furthermore, as mentioned in Sec.~\ref{nm2}, the total effect of the two laser beams is the summation of the effect of each beam, and thus
the effective Hamiltonian for the subspace of $^3$P$_2$ states ($F=3/2$) can be expressed as
\begin{eqnarray}
H_{\rm eff}=H_{\rm 1b}-i\frac{1}{2}H^\prime,
\end{eqnarray}
where the Hermitian term $H_{\rm 1b}$ is given by
\begin{eqnarray}
H_{\rm 1b}&=&\varepsilon_{1}(I_z,\omega_z)\!\sum_{\eta=\uparrow,\downarrow}|c,\eta\rangle\langle c,\eta| +\varepsilon_{3}(I_z,\omega_z)\!\sum_{\eta=\pm 3/2}|c,\eta\rangle\langle c,\eta|+\varepsilon_{1}(I_x,\omega_x)\!\sum_{\eta=\uparrow,\downarrow}|x,\eta\rangle\langle x,\eta| +\varepsilon_{3}(I_x,\omega_x)\!\sum_{\eta=\pm 3/2}|x,\eta\rangle\langle x,\eta|.\nonumber\\
\end{eqnarray}
with $\omega_{x(z)}$ and $I_{x(z)}$ being the angular frequency and intensity of the beam polarized along the $x$- ($z$-) direction, respectively, and the operator $H^\prime$ which describes the laser-induced atom loss is
\begin{eqnarray}
H^\prime&=&\lambda_{1}(I_z,\omega_z)\!\sum_{\eta=\uparrow,\downarrow}|c,\eta\rangle\langle c,\eta| +\lambda_{3}(I_z,\omega_z)\!\sum_{\eta=\pm 3/2}|c,\eta\rangle\langle c,\eta|+\lambda_{1}(I_x,\omega_x)\!\sum_{\eta=\uparrow,\downarrow}|x,\eta\rangle\langle x,\eta| +\lambda_{3}(I_x,\omega_x)\!\sum_{\eta=\pm 3/2}|x,\eta\rangle\langle x,\eta|.\nonumber\\
\end{eqnarray}
After re-choosing the zero-energy point, we can further express $H_{\rm 1b}$ as
\begin{eqnarray}
H_{\rm 1b}&=&\left[\Delta_{\rm AC}^{(z)}-\Delta_{\rm AC}^{(x)}/2\right]|c,3/2\rangle\langle c,3/2|+\frac{\sqrt{3}}4\Delta_{\rm AC}^{(x)}\Big[|c,\uparrow\rangle\langle c,3/2|+|c,3/2\rangle\langle c,\uparrow|\Big]\nonumber\\
&&+\left[\Delta_{\rm AC}^{(z)}-\Delta_{\rm AC}^{(x)}/2\right]|c,-3/2\rangle\langle c,-3/2|+\frac{\sqrt{3}}4\Delta_{\rm AC}^{(x)}\Big[|c,\downarrow\rangle\langle c,-3/2|+|c,-3/2\rangle\langle c,\downarrow|\Big],
\end{eqnarray}
with
\begin{eqnarray}
\Delta_{\rm AC}^{(z)}&=&\varepsilon_{3}(I_z,\omega_z)-\varepsilon_{1}(I_z,\omega_z);\\
\Delta_{\rm AC}^{(x)}&=&\varepsilon_{3}(I_x,\omega_x)-\varepsilon_{1}(I_x,\omega_x).
\end{eqnarray}
It is clear that the eigen states of $H_{\rm 1b}$ are the dressed states $|h(q),\uparrow\rangle$ and $|h(q),\downarrow\rangle$ defined in Eqs.~(\ref{hq1}-\ref{hq4}), with the coefficients $C$ and $C^\prime$ being given by
\begin{eqnarray}
C&=&\frac{1}{\sqrt{1+\frac{\left[1-2\Delta_{\rm AC}^{(z)}/\Delta_{\rm AC}^{(x)}+2\sqrt{1-\Delta_{\rm AC}^{(z)}/\Delta_{\rm AC}^{(x)}+\left(\Delta_{\rm AC}^{(z)}/\Delta_{\rm AC}^{(x)}\right)^2}\right]^2}{3}}};\label{expc}\\
\nonumber\\
\nonumber\\
C^\prime&=&\frac{\left\{1-2\Delta_{\rm AC}^{(z)}/\Delta_{\rm AC}^{(x)}+2\sqrt{1-\Delta_{\rm AC}^{(z)}/\Delta_{\rm AC}^{(x)}+\left(\Delta_{\rm AC}^{(z)}/\Delta_{\rm AC}^{(x)}\right)^2}\right\}/\sqrt{3}}{\sqrt{1+\frac{\left[1-2\Delta_{\rm AC}^{(z)}/\Delta_{\rm AC}^{(x)}+2\sqrt{1-\Delta_{\rm AC}^{(z)}/\Delta_{\rm AC}^{(x)}+\left(\Delta_{\rm AC}^{(z)}/\Delta_{\rm AC}^{(x)}\right)^2}\right]^2}{3}}}.\label{expcp}
\end{eqnarray}
Moreover, the eigen energies corresponding to  $|h,\uparrow\rangle$ and $|q,\uparrow\rangle$ are same as the one corresponding to  $|h,\downarrow\rangle$ and $|q,\downarrow\rangle$, respectively. The energy gap $\Delta_{hq}$ between the higher eigen states $|h,\uparrow\!(\downarrow)\rangle$ and the lower ones $|q,\uparrow\!(\downarrow)\rangle$ can be expressed as Eq.~\ref{dhqq} of Sec.~\ref{nm2}, i.e.,
\begin{eqnarray}
\Delta_{hq}=\sqrt{\left[\Delta_{\rm AC}^{(z)}-\Delta_{\rm AC}^{(x)}/2\right]^2+\frac34\Delta_{\rm AC}^{(x)2}}.
\end{eqnarray}
Finally, the total photon scattering rate $\Gamma_{\rm sc}$ of the laser beams can be estimated as
\begin{eqnarray}
\Gamma_{\rm sc}=\langle q,\downarrow|H^\prime|q,\downarrow\rangle=\langle q,\uparrow|H^\prime|q,\uparrow\rangle.
\end{eqnarray}

\subsection{The System of Method III}
\label{mm2}

In this section, we calculate the   effective Hamiltonian and the heating rate for the systems of method III,
and derive  the effective Rabi frequency $\Omega_{\rm eff}$ of the Raman coupling as the photon scattering rate $\Gamma_{\rm sc}$. We show the final results  in Eq.~(\ref{oeff22}) and Eq.~(\ref{gaa}).

As in Appendix~\ref{a1},
we take the states with magnetic quantum numbers $-1/2$ ($\uparrow$)
and $-3/2$
as an example.
Accordingly, we  consider the following $^3$P states:
\begin{eqnarray}
 |c,0\rangle&\equiv&|^3{\rm P}_0,1/2,-1/2\rangle
 =|^3{\rm P}_0,m_J=0\rangle_{ele}|-1/2\rangle_{nuc};
 \\
  \vspace{0.5cm} \nonumber\\
|c,\uparrow\rangle&\equiv&|^3{\rm P}_2,3/2,-1/2\rangle\nonumber\\
 &&=\sqrt{\frac 25}|^3{\rm P}_2,m_J=0\rangle_{ele}|-1/2\rangle_{nuc}-\sqrt{\frac 35}|^3{\rm P}_2,m_J=-1\rangle_{ele}|+
 1/2\rangle_{nuc};\\
 \vspace{0.5cm}\nonumber\\
|c,-3/2\rangle&\equiv&|^3{\rm P}_2,3/2,-3/2\rangle=|^3{\rm P}_2,m_J=-1\rangle_{ele}|-1/2\rangle_{nuc},
\end{eqnarray}
as well as the $^3{\rm S}_1$ and $3{\rm D}_{1,2,3}$ states $|f_{1,...,12}\rangle$ defined in Eqs.~(\ref{f1}-\ref{f12}). Notice that here definitions $|c,\uparrow\rangle$ and $|c,-3/2\rangle$ are the same as in Appendix~\ref{a1} and the main text. In this section we use the current notations just for simplicity.  As in Appendix~\ref{a1},
according to the selection rule, $|f_{1,2}\rangle$ can be coupled to the states $|c,0\rangle$ and $|c,\uparrow\rangle$ by the Raman beams, while $|f_{3,...,8}\rangle$ and $|f_{9,...,12}\rangle$ are coupled only to  $|c,\uparrow\rangle$ and $|c,-3/2\rangle$, respectively.

As shown in Fig.~\ref{scheme}(a) of the main text, we denote the two $\pi$-polarized Raman beams as $\alpha$ and $\beta$, with angular frequencies $\omega_\alpha$ and $\omega_\beta$, respectively ($\omega_\alpha>\omega_\beta$). Thus, in the Schr\"odinger picture  the Hamiltonian for our system can be expressed as ($\hbar=1$):
\begin{eqnarray}
H_S&=&\sum_{q=0,\uparrow,-\frac32}E_q|c,q\rangle\langle c, q|
+\sum_{j=1}^{12} \left(E_j-i\gamma_j/2\right)|f_j\rangle\langle f_j|
+
\sum_{q=0,\uparrow,-\frac32}\sum_{j=1}^{12}
\left\{\left[\Omega_{jq}^{(\alpha)}\cos\omega_\alpha t+\Omega_{jq}^{(\beta)}\cos\omega_\beta t\right]|f_j\rangle\langle c, q|+h.c.\right\}\nonumber\\
%\nonumber\\
%&&+\sum_{j=1}^8
%\left\{\left[\Omega_{jb}^{(\alpha)}\cos\omega_\alpha t+\Omega_{jb}^{(\beta)}\cos\omega_\beta t\right]|c_j\rangle{\color{red} \langle c,\uparrow|}+h.c.\right\},
\label{hs}
\end{eqnarray}
where $E_{q}$  is the energy of state $|c,q\rangle$ ($q=0,\uparrow,-\frac32$), while $E_{j}$  and $\gamma_j$
($j=1,...,8$)
are the energy and spontaneous emission rate
of excited state $|f_j\rangle$. According to the above definitions, we have
$
E_{j}=E_{j+4};\ \ \gamma_{j}=\gamma_{j+4},
$ $({j\leq 8})$.
In Eq.~(\ref{hs}) $\Omega_{jq}^{(\alpha(\beta))}$ ($q=0,\uparrow,-\frac32; j=1,...,12$) is the Rabi frequency of the coupling between states $|f_j\rangle$ and $|c,q\rangle$, which is induced by the
 beam $\alpha(\beta)$, and  can be further expressed as
 \begin{eqnarray}
 \Omega_{jq}^{(\zeta)}=\sqrt{\frac{2}{\epsilon_0c}}\langle f_j|D_z|c,q\rangle\sqrt{I_\zeta};\ \ \ (q=0,\uparrow,-\frac32;\  j=1,...,12;\ \zeta=\alpha,\beta),
 \end{eqnarray}
 where $\epsilon_0$ and $c$ are the vacuum dielectric constant and speed of light, respectively, $I_\zeta$ ($\zeta=\alpha,\beta$) is the intensity of the  beam $\zeta$, and  $\langle f_j|D_z{|c,q\rangle}$ is the matrix element of the atomic electric-dipole along the $z$-direction, with respect to the states $|f_j\rangle$ and ${|c,q\rangle}$, which is assumed to be real as before. As mentioned above,  the selection rules yield
$\langle f_{j}|D_z{|c,0\rangle}=\Omega_{ja}^{(\alpha,\beta)}=0$  for $j=3,...,12$,
$\langle f_{j}|D_z{|c,\uparrow\rangle}= \Omega_{jb}^{(\alpha,\beta)}=0$ for $j=9,...,12$, and
$\langle f_{j}|D_z{|c,-3/2\rangle}=\Omega_{jc}^{(\alpha,\beta)}=0$ for $j=1,...,8$.
 We derive the values of matrix elements $\langle f_{j}|D_z{|c,q\rangle}$ (${q=0,\uparrow,-\frac32}$, $j=1,...,12$) via the Wigner–Eckart theorem,  with the corresponding reduced matrix element given by Ref~\cite{Porsev1999} for $^{171}$Yb.

In Eq.~(\ref{hs}) we phenomenologically describe the spontaneous emission of the excited states via the non-Hermitian  term proportional to $i\gamma_j$ ($j=1,...,12$). In addition, for the seek of generality, we have considered the fact that each Raman beam can couple every $^3$P states {$|c,q\rangle$ $(q=0,\uparrow,-3/2)$}  to the excited states, and did not make the rotating-wave approximation.

 As shown in Sec.~\ref{m2}, we assume that both of the two beams are far-off resonant for the direct transitions from the $^3$P states to the excited states, while the angular frequency difference $\omega_\alpha-\omega_\beta$ is close to {$E_0-E_\uparrow$}.
 As a result, the  two-photon processes
\begin{eqnarray}
{|c,0\rangle}\xrightarrow{{\rm absorbing\ a\ photon\ with}\ \omega_\alpha}|f_{1,2}\rangle\xrightarrow{{\rm emitting\ a\ photon\ wiht}\ \omega_\beta}{ |c,\uparrow\rangle};\label{pro1}\\
{|c,0\rangle}\xrightarrow{{\rm emitting\ a\ photon\ with}\ \omega_\beta}|f_{1,2}\rangle\xrightarrow{{\rm absorbing\ a\ photon\ wiht}\ \omega_\alpha}{ |c,\uparrow\rangle},\label{pro2}
\end{eqnarray}
are near-resonant Raman processes. Here the process in Eq.~(\ref{pro2}) is caused by the anti-rotating-wave terms.
 In this case, we can adiabatically eliminate the excited states $|f_j\rangle$ ($j=1,...,12$) and derive the effective Hamiltonian for the $^3$P states {$|c,q\rangle$ $(q=0,\uparrow,-3/2)$}. In the rotating frame, this  effective Hamiltonian can be expressed as
 \begin{eqnarray}
H_{\rm eff}=
{\tilde \delta}{|c,0\rangle} {\langle c,0|}+\sum_{{q=0,\uparrow,-\frac32}}
\left[E^{(\rm AC)}_q-i\Gamma_q/2\right]{|c,q\rangle} {\langle c, q|}
+
\frac{\Omega_{\rm eff}}2\big({|c,0\rangle} {\langle c,\uparrow\!|}+{|c,\uparrow\rangle} {\langle c,0|}\big)-i\frac{1}{2}\Gamma_{0\uparrow}
\big({|c,0\rangle} {\langle c,\uparrow\!|}+{|c,\uparrow\rangle} {\langle c,0|}\big),\nonumber\\
\label{br}
\end{eqnarray}
where
  \begin{eqnarray}
{\tilde \delta}&=&(\omega_\alpha-\omega_\beta)-({E_{\uparrow}}-{E_{0}});\\
E^{(\rm AC)}_q&=&-\frac14\sum_{j=1}^{12}
\left[\frac{\Omega_{jq}^{(\alpha)2}}{E_j-E_q-\omega_\alpha}
+\frac{\Omega_{jq}^{(\alpha)2}}{E_j-E_q+\omega_\alpha}
+\frac{\Omega_{jq}^{(\beta)2}}{E_j-E_q-\omega_\beta}
+\frac{\Omega_{jq}^{(\beta)2}}{E_j-E_q+\omega_\beta}
\right];\nonumber\\
&&\hspace{10cm}({q=0,\uparrow,-\frac32});\\
\Omega_{\rm eff}&=&-\frac12\sum_{k=1,2}
\left[\frac{\Omega_{k\uparrow}^{(\beta)}\Omega_{k0}^{(\alpha)}}{E_k-{E_0}-\omega_\alpha}
+\frac{\Omega_{k\uparrow}^{(\alpha)}\Omega_{k0}^{(\beta)}}{E_k-{E_0}+\omega_\beta}
\right],\label{oeff22}
\end{eqnarray}
and
   \begin{eqnarray}
\Gamma_{q}&=&\frac14\sum_{j=1}^{12}\gamma_j
\left[\frac{\Omega_{jq}^{(\alpha)2}}{(E_j-E_q-\omega_\alpha)^2}
+\frac{\Omega_{jq}^{(\alpha)2}}{(E_j-E_q+\omega_\alpha)^2}
+\frac{\Omega_{jq}^{(\beta)2}}{(E_j-E_q-\omega_\beta)^2}
+\frac{\Omega_{jq}^{(\beta)2}}{(E_j-E_q+\omega_\beta)^2}
\right];\nonumber\\
&&\hspace{11.75cm}({q=0,\uparrow,-\frac32});\\
\Gamma_{0\uparrow}&=&\frac14\sum_{k=1,2}\gamma_k
\left[\frac{\Omega_{k\uparrow}^{(\alpha)}\Omega_{k0}^{(\beta)}}{(E_k-{E_0}-\omega_\alpha)^2}
+\frac{\Omega_{k\uparrow}^{(\beta)}\Omega_{k0}^{(\alpha)}}{(E_k-{E_0}+\omega_\beta)^2}
\right].
\label{brr}
\end{eqnarray}
Here ${\tilde \delta}$ is the two-photon detuning of the two Raman beams, {$E^{(\rm AC)}_q$ $(q=0,\uparrow,-3/2)$} are the AC-Stark shifts of states {$|c,q\rangle$}, respectively, which are induced by the Raman beams, and $\Omega_{\rm eff}$ is the effective Rabi frequency of the Raman transition between states ${ |c,0\rangle}$ and ${|c,\uparrow\rangle}$, as defined in our main text.
In addition, the heating effect given by the Raman beams is described by the terms with parameters $\Gamma_{q}$ {($q=0,\uparrow,-3/2$) and $\Gamma_{0\uparrow}$}. In the derivation of the results Eqs.~(\ref{br}-\ref{brr}) we have also used the fact $1/(g+i\xi/2)\approx 1/g-i\xi/(2g^2)$ with $g$ and $\xi$ being real numbers and $|g|\gg |\xi|$.

Furthermore, as mentioned in Sec.~\ref{1b}, in our scheme the frequencies of the Raman beams should be tuned to compensate the AC-Stark shift difference, i.e., the condition
\begin{eqnarray}
{\tilde \delta}+E^{(\rm AC)}_{0}=E^{(\rm AC)}_{\uparrow}
\end{eqnarray}
 is satisfied. Under this condition and after re-choosing the zero-energy point, we can further re-write
 the effective Hamiltonian $H_{\rm eff}$ as
  \begin{eqnarray}
H_{\rm eff}=H_{\rm 1b}-i\frac{1}{2}H^\prime,
\end{eqnarray}
with the Hermitian part
   \begin{eqnarray}
H_{\rm 1b}&=&\left(E^{(\rm AC)}_c-E^{(\rm AC)}_{\uparrow}\right){ |c,-3/2\rangle}{ \langle c,-3/2|}+
\frac{\Omega_{\rm eff}}2\big({ |c,0\rangle} { \langle c,\uparrow\!|}+{ |c,\uparrow\rangle} { \langle c,0|}\big),
\end{eqnarray}
describes the unitary evolution of the atom,
and
\begin{eqnarray}
H'=\sum_{{q=0,\uparrow,-\frac32}}
\Gamma_q{ |c,q\rangle} { \langle c, q|}
+
\Gamma_{0\uparrow}
\big({ |c,0\rangle} { \langle c,\uparrow\!|}+{ |c,\uparrow\rangle} { \langle c,0|}\big)
\end{eqnarray}
describes the heating effect. The eigen energies of $H_{\rm 1b}$, which are mentioned in Sec.~\ref{1b}, are just given by
\begin{eqnarray}
{\cal E}_d=-\frac{\Omega_{\rm eff}}2;\hspace{0.6cm} {\cal E}_u=\frac{\Omega_{\rm eff}}2;\hspace{0.6cm}
{\cal E}_c=E^{(\rm AC)}_{-3/2}-E^{(\rm AC)}_\uparrow.
\end{eqnarray}
These expressions yield the results in Eqs.~(\ref{eeu},\ref{1beg}).

We further estimate the photon scattering rate $\Gamma_{\rm sc}$ of the Raman beams as
\begin{eqnarray}
\Gamma_{\rm sc}=\langle d,\uparrow\!|H^\prime|d,\uparrow\rangle,\label{gaa}
\end{eqnarray}
with the lower dressed state $|d,\uparrow\rangle$ being defined in Eq.~(\ref{sd}) and being able to be expressed as
$
|d,\uparrow\rangle=({ |c,0\rangle}+{ |c,-3/2\rangle})/\sqrt{2},
$
with the notations of the current Appendix. The calculations for Fig.~\ref{heating} and related parts in Sec.~\ref{1b} are based on Eq.~(\ref{oeff22}) and Eq.~(\ref{gaa}).

\section{Inter-Atomic Interaction}
\label{int}

In this appendix, we show the models of inter-atomic interaction  used in this work. As mentioned in the main text, we label the two atoms as 1 and 2 and use ${\bf r}$ to denote the relative position of these two atoms. Then the inter-atomic interaction potential is  an ${\bf r}$-dependent operator for the Hilbert space ${{\mathscr H}_{\rm internal}}$ of the two-atom internal state. In addition, the space  ${{\mathscr H}_{\rm internal}}$ can be further factorized to
\begin{eqnarray}
{\mathscr H}_{\rm internal}={\mathscr H}_{e1}\otimes{{\mathscr H}_{e2}}\otimes{{\mathscr H}_{n1}}\otimes{{\mathscr H}_{n2}},
\end{eqnarray}
with ${\mathscr H}_{ei}$ and ${\mathscr H}_{ni}$ ($i=1,2$) being the Hilbert space for the outer-shell electrons and atomic core of atom $i$, respectively. In this appendix, we use the notation $|\rangle$ to denote the states in ${\mathscr H}_{\rm internal}$,
$|\rangle_{e}$ for the states in ${\mathscr H}_{e1}\otimes{{\mathscr H}_{e2}}$,
$|\rangle_{ei}$ and $|\rangle_{ni}$ ($i=1,2$) for the states in ${\mathscr H}_{ei}$ and ${\mathscr H}_{ni}$, respectively, and $|\rangle_{i}$  ($i=1,2$) for the states in ${\mathscr H}_{ei}\otimes{\mathscr H}_{ni}$.

\subsection{Inter-Atomic Interaction of Methods I and II}
\label{appb1}

The inter-atomic interaction ${\hat V}^{(2)}({\bf r})$ for the systems of method I and method II, which are studied in Sec.~\ref{ss1} and Sec.~\ref{nm2}, respectively, can be expressed as
\begin{eqnarray}
{\hat V}^{(2)}({\bf r})={\hat {\cal P}}{\hat V}^{\rm bare}_2({\bf r}){\hat {\cal P}},
\end{eqnarray}
where ${\hat V}^{\rm bare}_2({\bf r})$ is the interaction potential
between a $^1$S$_0$-atom and a $^3$P$_2$-atom with arbitrary atomic spin  $F$, and
${\hat {\cal P}}$ is the projection operator for the states where the $^3$P$_2$-atom is in the subspace with $F=3/2$, i.e.,
\begin{eqnarray}
{\hat {\cal P}}&=&\sum_{m_F={-3/2}}^{+3/2}|^3{\rm P}_2,3/2,m_F\rangle_1\langle ^3{\rm P}_2,3/2,m_F|\otimes|^1{\rm S}_0\rangle_{e2}\langle ^1{\rm S}_0|\otimes{\hat I}_{n2}\nonumber\\
&&+\sum_{m_F={-3/2}}^{+3/2}|^1{\rm S}_0\rangle_{e1}\langle ^1{\rm S}_0|\otimes{\hat I}_{n1}\otimes|^3{\rm P}_2,3/2,m_F\rangle_2\langle ^3{\rm P}_2,3/2,m_F|,\label{pex}
\end{eqnarray}
with ${\hat I}_{nj}$ $(j=1,2)$ being the identity operator in the space of ${\mathscr H}_{ni}$.
Moreover, according to the Born-Oppenheimer approximation, the inter-atomic interaction is determined by the energy of electronic states for fixed positions of the two atomic cores (i.e., fixed ${\bf r}$). Based on this principle, we express the bare $^1$S$_0$-$^3$P$_2$ interaction ${\hat V}^{\rm bare}_2({\bf r})$ as
\begin{eqnarray}
{\hat V}^{\rm bare}_2({\bf r})&=&
\sum_{j=0,\pm 1,\pm 2}
\Big[V_{ j,+}(r)|\psi_{j,+}({\hat {\bf r}})\rangle_{e}\langle \psi_{j,+}({\hat {\bf r}})|
+V_{ j,-}(r)|\psi_{j,-}({\hat {\bf r}})\rangle_{e}\langle \psi_{j,-}({\hat {\bf r}})|
\Big]\otimes {\hat I}_{n1}\otimes {\hat I}_{n2},\label{v2}
\end{eqnarray}
%\begin{eqnarray}
%{\hat V}^{\rm bare}_2({\bf r})&=&
%\sum_{j=+,-}
%\Big[V_{\Sigma j}(r)|\psi_{\Sigma j}({\hat {\bf r}})\rangle_{e}\langle \psi_{\Sigma j}({\hat {\bf r}})|
%+V_{\Pi j}(r){\hat {\cal P}}_{\Pi j}({\hat {\bf r}})
%+V_{\Delta j}(r){\hat {\cal P}}_{\Delta j}({\hat {\bf r}})
%\Big]\otimes {\hat I}_{n1}\otimes {\hat I}_{n2},\label{v2}
%\end{eqnarray}
where
$
{\hat {\bf r}}={\bf r}/r
$
is the unit vector along the direction of ${\bf r}$, and the
${\hat {\bf r}}$-dependent electronic state
$|\psi_{j \pm}({\hat {\bf r}})\rangle_{e}$ ($j=0,\pm1,\pm 2$) is the state in which the  $^3$P$_2$-atom is in the electronic state with $J_{\hat{\bf r}}=j$. Here $J_{\hat{\bf r}}$ is the electronic total angular momentum on the interatomic axis (i.e., $J_{\hat{\bf r}}=({\bf L}+{\bf S})\cdot {\hat {\bf r}}$, with ${\bf L}$ and ${\bf S}$ being the orbital angular momentum and spin of the outer-shell electrons of the $^3$P$_2$-atom). Explicitly, we have
\begin{eqnarray}
|\psi_{j \pm}({\hat {\bf r}})\rangle_{e}=\frac{1}{\sqrt{2}}\Big[|^1{\rm S}_0\rangle_{e1}|^3{\rm P}_2,J_{\hat{\bf r}}=j\rangle_{e2}\pm|^3{\rm P}_2,J_{\hat{\bf r}}=j\rangle_{e1}|^1{\rm S}_0\rangle_{e2}\Big],\ \ \ (j=0,\pm 1, \pm 2).
\end{eqnarray}
In addition, in Eq. (\ref{v2}) $V_{ j,\pm}(r)$ $(j=0,\pm 1, \pm 2)$ is the interaction potential corresponding to the electronic state $|\psi_{j \pm}({\hat {\bf r}})\rangle_{e}$. Due to the  symmetry of the electrons under the transformation $J_{\hat{\bf r}}\rightarrow -J_{\hat{\bf r}}$, we have $V_{ -1,\pm}(r)=V_{ 1,\pm}(r)$ and $V_{ -2,\pm}(r)=V_{ 2,\pm}(r)$. Therefore, in this model the $^1$S$_0$-$^3$P$_{2}$ interaction
${\hat V}^{(2)}({\bf r})$ is determined by the  six potential curves:
\begin{eqnarray}
\{V_{0,\pm}(r),V_{1,\pm}(r),V_{2,\pm}(r)\}.
\end{eqnarray}
In Sec.~\ref{model1}
they are denoted as  $V^{(2)}_{1,...,6}(r)$, explicitly, we have
\begin{eqnarray}
V^{(2)}_{1}(r)&\equiv& V_{0,+}(r);\hspace{1cm}
V^{(2)}_{2}(r)\equiv V_{0,-}(r);\\
V^{(2)}_{3}(r)&\equiv& V_{1,+}(r);\hspace{1cm}
V^{(2)}_{4}(r)\equiv V_{1,-}(r);\\
V^{(2)}_{5}(r)&\equiv& V_{2,+}(r);\hspace{1cm}
V^{(2)}_{6}(r)\equiv V_{2,-}(r).
\end{eqnarray}
In addition, the operators ${\hat D}_{1,...,6}({\hat {\bf r}})$ in Sec.~\ref{model1} are just the ones proportional to $V^{(2)}_{1,...,6}(r)$ in the expression of ${\hat {\cal P}}{\hat V}^{\rm bare}_2({\bf r}){\hat {\cal P}}$, i.e.,
\begin{eqnarray}
{\hat D}_{1}({\hat {\bf r}})&=&{\hat {\cal P}}|\psi_{0,+}({\hat {\bf r}})\rangle_{e}\langle \psi_{0,+}({\hat {\bf r}})|{\hat {\cal P}};\\
{\hat D}_{2}({\hat {\bf r}})&=&{\hat {\cal P}}|\psi_{0,-}({\hat {\bf r}})\rangle_{e}\langle \psi_{0,-}({\hat {\bf r}})|{\hat {\cal P}};\\
{\hat D}_{3}({\hat {\bf r}})&=&{\hat {\cal P}}\big[|\psi_{1,+}({\hat {\bf r}})\rangle_{e}\langle \psi_{1,+}({\hat {\bf r}})|+
|\psi_{-1,+}({\hat {\bf r}})\rangle_{e}\langle \psi_{-1,+}({\hat {\bf r}})|
\big]{\hat {\cal P}};\\
{\hat D}_{4}({\hat {\bf r}})&=&{\hat {\cal P}}\big[|\psi_{1,-}({\hat {\bf r}})\rangle_{e}\langle \psi_{1,-}({\hat {\bf r}})|+
|\psi_{-1,-}({\hat {\bf r}})\rangle_{e}\langle \psi_{-1,-}({\hat {\bf r}})|
\big]{\hat {\cal P}};\\
{\hat D}_{5}({\hat {\bf r}})&=&{\hat {\cal P}}\big[|\psi_{2,+}({\hat {\bf r}})\rangle_{e}\langle \psi_{2,+}({\hat {\bf r}})|+
|\psi_{-2,+}({\hat {\bf r}})\rangle_{e}\langle \psi_{-2,+}({\hat {\bf r}})|
\big]{\hat {\cal P}};\\
{\hat D}_{6}({\hat {\bf r}})&=&{\hat {\cal P}}\big[|\psi_{2,-}({\hat {\bf r}})\rangle_{e}\langle \psi_{2,-}({\hat {\bf r}})|+
|\psi_{-2,-}({\hat {\bf r}})\rangle_{e}\langle \psi_{-2,-}({\hat {\bf r}})|
\big]{\hat {\cal P}}.
\end{eqnarray}

Finally, in two-body calculation, we require to express the eigen state $|^3{\rm P}_2,J_{\hat{\bf r}}=j\rangle_{e\xi}$ ($\xi=1,2$) of $J_{\hat{\bf r}}$ in terms of the eigen states of $J_{z}\equiv ({\bf L}+{\bf S})\cdot {\hat {\bf e}}_z$, where ${\hat {\bf e}}_z$ is the unit vector along the $z$-axis of the lab frame. This can be done via the relation
\begin{eqnarray}
|^3{\rm P}_2,J_{\hat{\bf r}}=j\rangle_{e\xi}=\sum_{\eta=0,\pm1,\pm 2}
|^3{\rm P}_2,J_{z}=\eta\rangle_{e\xi}D^{(2)}_{\eta,j}(\lambda_1,\lambda_2,0),\ \ (\xi=1,2).
\end{eqnarray}
Here $D^{(2)}_{\eta,j}(\lambda_1,\lambda_2,0)$ is the Winger D-function
\cite{Rose1957}, with $\lambda_1$ and $\lambda_2$ being the  azimuthal angle and polar angle of ${\hat {\bf r}}$ in the lab frame, i.e., ${\hat {\bf r}}=\cos\lambda_2{\hat {\bf e}}_z+\sin\lambda_2\cos\lambda_1{\hat {\bf e}}_x+\sin\lambda_2\sin\lambda_1{\hat {\bf e}}_y$, where ${\hat {\bf e}}_{x(y)}$ is the unit vectors along the $x(y)$-axis of the lab frame.

\subsection{Inter-Atomic Interaction of Method III}
\label{appb2}

As shown in  Sec.~\ref{ss},
for the systems of method III,   the inter-atomic interaction is
\begin{eqnarray}
 {\hat V}({\bf r})\equiv{\hat V}^{(0)}(r)+{\hat V}^{(2)}({\bf r}),
 \end{eqnarray}
 where ${\hat V}^{(2)}({\bf r})$  is the one derived in the above subsection
 and ${\hat V}^{(0)}(r)$ is the
 interaction between an atom in $^1$S$_0$ state and another atom in $^3$P$_{0}$ state. As shown in previous works Ref~\cite{Cappellini2014,Scazza2014}, the potential ${\hat V}^{(0)}(r)$ can be expressed as
\begin{eqnarray}
{\hat V}^{(0)}(r)=\Big[V_+(r)|\psi_+\rangle_{e}\langle \psi_+|+V_-(r)|\psi_-\rangle_{e}\langle \psi_-|\Big]\otimes {\hat I}_{n1}\otimes {\hat I}_{n2}.
\end{eqnarray}
Here $ {\hat I}_{ni}$ ($i=1,2$) is the identical operator of ${\mathscr H}_{ni}$, $V_{\pm}(r)$ is the interaction potential curve corresponding to the electronic states $|\psi_\pm\rangle_{e}\equiv \frac{1}{\sqrt{2}}\Big[|^1{\rm S}_0\rangle_{e1}|^3{\rm P}_0\rangle_{e2}\pm|^3{\rm P}_0\rangle_{e1}|^1{\rm S}_0\rangle_{e2}\Big]$.

\section{Parameters for the Calculations with Multi-Channel Square-Well Models}
\label{apppara}

In the following tables, we show the depth $U_{1,...,6}^{(2)}$ and $U_{\pm}^{(0)}$  used in our calculations with multi-channel square-well models  in Sec.~\ref{model1}, Sec.~\ref{nm2}, and  Sec.~\ref{model2}, for methods I, II, and III, respectively. Here we also show the $s$-wave  scattering length $a_{1,...,6}^{(2)}$ and $a_{\pm}^{(0)}$ corresponding to the single-channel square-well potentials $-U_{1,...,6}^{(2)}\theta(b-r)$ and $-U_{\pm}^{(0)}\theta(b-r)$, respectively, with $\theta(x)$ being the step function and $b$ is same as the width of our multi-channel models, i.e., $b=85a_0$ for methods I and II and $b=81a_0$ for method III.
In the tables, the unit of $U_{1,...,6}^{(2)}$ and $U_{\pm}^{(0)}$ is $(2\pi)$MHz ($\hbar=1$) and the unit of $a_{1,...,6}^{(2)}$ and $a_{\pm}^{(0)}$ is the Bohr radius $a_0$.

\begin{table}[h]
\caption{U ($2\pi$MHz) and $a$ $(a_0)$ of Method I}
\centering
\def\temptablewidth{1\textwidth}
{\rule{\temptablewidth}{1pt}}  %žùŸÝÊ¹ÓÃÇé¿öÁé»îÉèÖÃ£¬ÏßµÄŽÖÏž
\begin{tabular*}{\temptablewidth}{@{\extracolsep{\fill}}ccccccccccccccc}
\hline
& &$U^{(2)}_1$&$a^{(2)}_1$&$U^{(2)}_2$&$a^{(2)}_2$&$U^{(2)}_3$&$a^{(2)}_3$&$U^{(2)}_4$&$a^{(2)}_4$&$U^{(2)}_5$&$a^{(2)}_5$&$U^{(2)}_6$&$a^{(2)}_6$&\\
\hline
&case (1)& 2862.99 & 85.35 & 2796.73 & 86.47 & 2228.74 & 87.43 & 2693.10 & 90.11 & 2194.22 & 88.77 & 2629.19 & 105.27 &\\
\hline
&case (2)& 2809.38 & 86.23 & 2457.57 & 82.46 & 2757.31 & 87.38 & 2692.93 & 90.12 & 2714.58 & 88.89 & 2729.90 & 88.25 &\\
\hline
\end{tabular*}
{\rule{\temptablewidth}{1pt}}
\end{table}

\begin{table}[h]
\caption{U ($2\pi$MHz) and $a$ $(a_0)$ of Method II}
\centering
\def\temptablewidth{1\textwidth}
{\rule{\temptablewidth}{1pt}}  %žùŸÝÊ¹ÓÃÇé¿öÁé»îÉèÖÃ£¬ÏßµÄŽÖÏž
\begin{tabular*}{\temptablewidth}{@{\extracolsep{\fill}}ccccccccccccccc}
\hline
& &$U^{(2)}_1$&$a^{(2)}_1$&$U^{(2)}_2$&$a^{(2)}_2$&$U^{(2)}_3$&$a^{(2)}_3$&$U^{(2)}_4$&$a^{(2)}_4$&$U^{(2)}_5$&$a^{(2)}_5$&$U^{(2)}_6$&$a^{(2)}_6$&\\
\hline
&case (1)& 2267.92 & 86.39 & 2336.87 & 85.02 & 2719.33 & 88.68 & 2891.00 & 84.93 & 2726.08 & 88.40 & 2268.16 & 86.38 &\\
\hline
&case (2)& 2581.53 & 63.79 & 2311.49 & 85.49 & 2112.22 & 103.26 & 2196.97 & 88.64 & 2905.45 & 84.72 & 2729.59 & 88.26 &\\
\hline
\end{tabular*}
{\rule{\temptablewidth}{1pt}}
\end{table}

\begin{table}[!h]
\caption{U ($2\pi$MHz) and $a$ $(a_0)$ of Method III}
\centering
\def\temptablewidth{1\textwidth}
{\rule{\temptablewidth}{1pt}}  %žùŸÝÊ¹ÓÃÇé¿öÁé»îÉèÖÃ£¬ÏßµÄŽÖÏž
\begin{tabular*}{\temptablewidth}{@{\extracolsep{\fill}}ccccccccccccccccccc}
\hline
& & $U^{(0)}_+$ & $a^{(0)}_+$ & $U^{(0)}_-$ & $a^{(0)}_-$ & $U^{(2)}_1$ & $a^{(2)}_1$ & $U^{(2)}_2$ & $a^{(2)}_2$ & $U^{(2)}_3$ & $a^{(2)}_3$ & $U^{(2)}_4$ &  $a^{(2)}_4$ &$U^{(2)}_5$ &$a^{(2)}_5$ & $U^{(2)}_6$ & $a^{(2)}_6$ & \\
\hline
&case (1)& 647.08 & 232.00 & 391.14 & 372.00 & 2216.60 & 74.89 & 2300.85 & 198.01 & 2273.99 & 57.79 & 2718.70 & 78.24 & 2289.44 & 5.97 & 2682.05 & 79.14 &\\
\hline
&case (2)& 392.08 & 232.00 & 206.11 & 372.00 & 2416.42 & 84.59 & 2203.85 & 75.88 & 2554.87 & 81.31 & 2136.36 & 78.74 & 2166.35 & 77.76 & 2191.94 & 76.60 &\\
\hline
\end{tabular*}
{\rule{\temptablewidth}{1pt}}
\end{table}

{
\section{Conditions and Interaction Parameters for $A_0=0$}
\label{appd}

In the following tables, we show the values of the control parameters $\Delta_{\rm AC}$, $\Delta_{hq}$ and $\Omega_{\rm eff}$ under which we have  $A_0=0$ for the  cases in Fig.~\ref{sw1}, Fig.~\ref{rsw2} and Fig.~\ref{sw2}, respectively.
We also show the interaction parameters $A_{x,y,z}$ and the scattering lengths $a_{\pm, f}$  (or $a_{\pm, p\pm}$) at $A_0=0$ for these cases, in the unit of the width $b$ of the corresponding square-well models.

\begin{table}[!h]
\caption{Conditions and interaction parameters for $A_0=0$ in the cases of FIG.~\ref{sw1}.}
\centering
\def\temptablewidth{0.7\textwidth}
{\rule{\temptablewidth}{1pt}}
\begin{tabular*}{\temptablewidth}{@{\extracolsep{\fill}}cccccccc}
\hline
                           & $\Delta_{\rm{AC}}/((2\pi) \rm{MHz})$ & $a_+/b$ & $a_-/b$ & $a_f/b$ & $A_x/b=A_y/b$ & $A_z/b$ &\\ \hline
 \multirow{2}{*}{case (1)} & 12.44              & 1.06  & 2.69  & -1.84 & 0.82      & -3.72 &\\ \cline{2-8}
                           & 15.40              & 1.06  & -2.01 & 0.48  & -1.53     & 0.96  &\\  \hline
 \multirow{2}{*}{case (2)} & 3.29               & 3.68  & 1.08  & -2.22 & -1.30     & -4.60  &\\  \cline{2-8}
                           & 4.57               & -2.69 & 1.08  & 0.81  & 1.88      & 1.62   &\\
\hline
\end{tabular*}
{\rule{\temptablewidth}{1pt}}
\end{table}

\begin{table}[!h]
\caption{Conditions and interaction parameters of for $A_0=0$ in the cases of FIG.~\ref{rsw2}.}
\centering
\def\temptablewidth{0.7\textwidth}
{\rule{\temptablewidth}{1pt}}
\begin{tabular*}{\temptablewidth}{@{\extracolsep{\fill}}cccccccccc}
\hline
                          & $\Delta_{hq}/((2\pi) \rm{MHz})$ & $a_+/b$ & $a_-/b$ & $a_{p+}/b$ & $a_{p-}/b$ & $A_x/b$ & $A_y/b$ & $A_z/b$ &\\ \hline
\multirow{2}{*}{case (1)} & 8.20               & -3.39 & 1.21  & 1.20     & 1.69     & 2.54  & 2.05  & 2.53  &\\ \cline{2-10}
                          & 11.06              & 1.15  & 1.25  & 1.22     & -3.59    & -2.36 & 2.46  & -2.38 &\\ \hline
\multirow{2}{*}{case (2)} & 2.73               & 2.10  & 1.03  & -6.26    & 2.03     & 3.61  & -4.67 & -3.68 &\\ \cline{2-10}
                          & 4.71               & -4.72 & 1.03  & 0.94     & 2.78     & 3.80  & 1.96  & 3.71  &\\
\hline
\end{tabular*}
{\rule{\temptablewidth}{1pt}}
\end{table}

\begin{table}[!h]
\caption{Conditions and interaction parameters of for $A_0=0$ in the cases of FIG.~\ref{sw2}.}
\centering
\def\temptablewidth{0.7\textwidth}
{\rule{\temptablewidth}{1pt}}
\begin{tabular*}{\temptablewidth}{@{\extracolsep{\fill}}cccccccc}
\hline
         & $\Omega_{\rm{eff}}/((2\pi) \rm{MHz})$ & $a_+/b$ & $a_-/b$ & $a_{f}/b$ & $A_x/b=A_y/b$ & $A_z/b$ &\\  \hline
case (1) & 1.39                & 6.04  & -2.01 & -2.05   & -4.03     & -4.06 &\\  \hline
case (2) & 1.66                & 10.87 & -4.28 & -3.32   & -7.58     & -6.61 &\\
\hline
\end{tabular*}
{\rule{\temptablewidth}{1pt}}
\end{table}

}

\end{widetext}

%Here $\Sigma$, $\Pi$ and $\Delta$ correspond to the electronic states with total angular momentum $\Lambda$ along the interatomic axis satisfying $|\Lambda|=0,1,2$, respectively, and the label $\pm$ indicates that the electronic state is symmetric/anti-symmetric with respect to the permutation of the two atoms.

%which
%is comparable with the depth of the optical trapping potential.

\bibliography{references}

\begin{thebibliography}{48}
\bibitem [{foo()}]{footnote1}%
  \BibitemOpen
  \href@noop {} {}\bibinfo {note} {For a van der Waals potential $-C_6/r$,
  $\beta_6$ and $E_{\rm vdW}$ are defined as $E_{\rm
  vdW}=2\hbar^2/(\mu\beta_6^2)$, respecitvely, with $\mu$ being the reduced
  mass}\BibitemShut {NoStop}%
\bibitem [{3pa()}]{3page}%
  \BibitemOpen
  \href@noop {} {}\bibinfo {note} {{ In this work, the detuning of a laser beam
  with respect to an atomic transition is defined as $\omega_{\rm
  A}-\omega_{\rm L}$, with $\omega_{\rm A}$ and $\omega_{\rm L}$ being the
  angular frequency of the atomic transition and the laser, respectively.
  }}\BibitemShut {NoStop}%
\bibitem [{two()}]{two6page}%
  \BibitemOpen
  \href@noop {} {}\bibinfo {note} {{ Following the previous studies for the
  magnetic Feshbach resonances of alkaline atoms with square-well models, we
  choose the width of the square well as $b={\bar a}$, where ${\bar a}\equiv
  2\pi/\Gamma[1/4]^2\beta_6$ is the ``van der Waals length
  scale"\cite{RMPChin}. For the $^1$S$_0$-$^3$P$_0$ interaction of two
  $^{171}$Yb atoms, it was found that $\beta_6\approx 168a_0$ \cite{Yejun2014}
  and thus ${\bar a}\approx 81a_0$. However, for the $^1$S$_0$-$^3$P$_2$
  interaction of two $^{171}$Yb atoms, we do not know the value of $\beta_6$ or
  ${\bar a}$. We estimate ${\bar a}$ for this case to be $85a_0$, and thus
  choose $b=85a_0$ for methods I and II. In addition, in the system of method
  III, one atom is in a $^3$P$_0$-$^3$P$_2$ dressed state and the other one is
  in the $^1$S$_0$. The inter-atomic interaction potential is {\it different}
  with both the $^1$S$_0$-$^3$P$_0$ potential and the $^1$S$_0$-$^3$P$_2$
  potential, and the value of ${\bar a}$ is unknown. In the corresponding
  calculations, for simplicity, we just choose $b=81a_0$. Here we emphasis that
  the qualitative results ({\it i.e.}, the order of magnitude of the laser
  intensities with which the resonances can occur ) of our calculations are the
  same when $b$ is around the values used in our manuscript. }}\BibitemShut
  {NoStop}%
\end{thebibliography}

\end{document}